\documentclass[12pt,epsfig]{article}
\usepackage{epsfig}
\usepackage{graphicx}
\begin{document}


\newcommand{\beq}{\begin{equation}}
\newcommand{\eeq}{\end{equation}}
\newcommand{\bea}{\begin{eqnarray}}
\newcommand{\eea}{\end{eqnarray}}
\newcommand{\beqn}{\begin{eqnarray}}
\newcommand{\eeqn}{\end{eqnarray}}
\newcommand{\beas}{\begin{eqnarray*}}
\newcommand{\eeas}{\end{eqnarray*}}
\newcommand{\defi}{\stackrel{\rm def}{=}}
\newcommand{\non}{\nonumber}
\newcommand{\bquo}{\begin{quote}}
\newcommand{\enqu}{\end{quote}}
\newcommand{\p}{\partial}


\def\de{\partial}
\def\Tr{ \hbox{\rm Tr}}
\def\const{\hbox {\rm const.}}
\def\o{\over}
\def\im{\hbox{\rm Im}}
\def\re{\hbox{\rm Re}}
\def\bra{\langle}\def\ket{\rangle}
\def\Arg{\hbox {\rm Arg}}
\def\Re{\hbox {\rm Re}}
\def\Im{\hbox {\rm Im}}
\def\diag{\hbox{\rm diag}}

\def\stroke{\vrule height8pt width0.4pt depth-0.1pt}
\def\topfleck{\vrule height8pt width0.5pt depth-5.9pt}
\def\botfleck{\vrule height2pt width0.5pt depth0.1pt}
\def\Zmath{\vcenter{\hbox{\numbers\rlap{\rlap{Z}\kern 0.8pt\topfleck}\kern
2.2pt\rlap Z\kern 6pt\botfleck\kern 1pt}}}
\def\Qmath{\vcenter{\hbox{\upright\rlap{\rlap{Q}\kern
3.8pt\stroke}\phantom{Q}}}}
\def\Nmath{\vcenter{\hbox{\upright\rlap{I}\kern 1.7pt N}}}
\def\Cmath{\vcenter{\hbox{\upright\rlap{\rlap{C}\kern
3.8pt\stroke}\phantom{C}}}}
\def\Rmath{\vcenter{\hbox{\upright\rlap{I}\kern 1.7pt R}}}
\def\Z{\ifmmode\Zmath\else$\Zmath$\fi}
\def\Q{\ifmmode\Qmath\else$\Qmath$\fi}
\def\N{\ifmmode\Nmath\else$\Nmath$\fi}
\def\C{\ifmmode\Cmath\else$\Cmath$\fi}
\def\R{\ifmmode\Rmath\else$\Rmath$\fi}


\def\QATOPD#1#2#3#4{{#3 \atopwithdelims#1#2 #4}}
\def\stackunder#1#2{\mathrel{\mathop{#2}\limits_{#1}}}
\def\stackreb#1#2{\mathrel{\mathop{#2}\limits_{#1}}}
\def\Tr{{\rm Tr}}
\def\res{{\rm res}}
\def\Bf#1{\mbox{\boldmath $#1$}}
\def\balpha{{\Bf\alpha}}
\def\bbeta{{\Bf\beta}}
\def\bgamma{{\Bf\gamma}}
\def\bnu{{\Bf\nu}}
\def\bmu{{\Bf\mu}}
\def\bphi{{\Bf\phi}}
\def\bPhi{{\Bf\Phi}}
\def\bomega{{\Bf\omega}}
\def\blambda{{\Bf\lambda}}
\def\brho{{\Bf\rho}}
\def\bsigma{{\bfit\sigma}}
\def\bxi{{\Bf\xi}}
\def\bbeta{{\Bf\eta}}
\def\d{\partial}
\def\der#1#2{\frac{\d{#1}}{\d{#2}}}
\def\Im{{\rm Im}}
\def\Re{{\rm Re}}
\def\rank{{\rm rank}}
\def\diag{{\rm diag}}
\def\2{{1\over 2}}
\def\ntwo{${\cal N}=2\;$}
\def\4N{${\cal N}=4$}
\def\none{${\cal N}=1\;$}
\def\x{\stackrel{\otimes}{,}}
\def\beq{\begin{equation}}
\def\eeq{\end{equation}}
\def\beqn{\begin{eqnarray}}
\def\eeqn{\end{eqnarray}}
\def\ba{\beq\new\begin{array}{c}}
\def\ea{\end{array}\eeq}
\def\be{\ba}
\def\ee{\ea}
\def\stackreb#1#2{\mathrel{\mathop{#2}\limits_{#1}}}

\def\baselinestretch{1.0}

\begin{titlepage}

\begin{flushright}
FTPI-MINN-06-17\\
UMN-TH-2504-06\\
\end{flushright}

\vspace{1cm}

\begin{center}

{\Large \bf Domain Lines as
  Fractional Strings}
\end{center}

\vspace{0.5cm}

\begin{center}
{{\bf R.~Auzzi}$^{\,a}$, {\bf M.~Shifman}$^{\,a}$ and  {\bf A.~Yung}$^{\,a,b,c}$}
\end {center}
\begin{center}

$^a${\it  William I. Fine Theoretical Physics Institute,
University of Minnesota,
Minneapolis, MN 55455, USA}\\
$^b${\it Petersburg Nuclear Physics Institute, Gatchina, St. Petersburg
188300, Russia}\\
$^c${\it Institute of   Theoretical and Experimental Physics,
Moscow  117250, Russia}\\

\end{center}

\vspace{3mm}

\begin{abstract}

We consider $\mathcal{N}=2$
supersymmetric quantum electrodynamics  (SQED)
with 2 flavors, the Fayet--Iliopoulos parameter,  and a
mass term $\beta$ which breaks the 
extended supersymmetry down to $\mathcal{N}=1$. The bulk theory has two vacua;
 at $\beta=0$
the BPS-saturated domain wall interpolating between them has a moduli space
parameterized by  a $U(1)$ phase $\sigma$ which 
can be promoted to a scalar field in the
effective low-energy theory on the wall world-volume.
At small nonvanishing $\beta$ this field gets a sine-Gordon potential.
As a result,
only two discrete degenerate BPS domain walls survive.
We find an explicit solitonic solution for domain lines ---
string-like objects  living on the surface
of the domain wall which separate wall I from wall II.
The domain line  
is seen as a BPS kink in the world-volume effective theory. 
 We expect that the wall with the domain line on it 
 saturates both the $\{1,0\}$ and the $\{\frac{1}{2},\frac{1}{2}\}$
  central charges of the bulk theory.
The domain line carries a magnetic flux
which is exactly $\frac{1}{2}$ of the flux carried by the flux tube living
in the bulk on each side of the wall. Thus, the domain lines on the wall  confine
charges living on the wall,  
resembling Polyakov's three-dimensional confinement.

\end{abstract}

\end{titlepage}

\section{Introduction}
\label{intro}

Supersymmetric   gauge theories were shown to support a large variety
of extended critical (i.e. BPS-saturated) objects exhibiting various nontrivial dynamical patterns. In particular,
gauge theories with extended $\mathcal{N}=2$
supersymmetry (SUSY) present a unique  framework in which many strong-coupling problems
can be modeled and  addressed in a quantitative way. The most
famous example of this type is the  Seiberg--Witten solution \cite{SW1,SW2}.

In the last few years we witnessed extensive explorations of various $\mathcal{N}=2$
models at weak coupling.
Solitonic BPS-saturated objects of novel types were found and studied,
such as non-Abelian flux tubes (strings) \cite{HT,ABEKY},
 domain-wall junctions  \cite{junct,GS,StV,jjunct,jwallweb}, domain-wall-string junctions
(boojums) \cite{SY-abw,j-boojums,SY-naw,rev1,ASY-boojums}, trapped monopoles
\cite{SY-mon,HT2,012} and so on, for  recent reviews see
\cite{rev1,rev2,Jrev}. Instantons inside domain walls in five dimensions
have been discussed recently in Ref. \cite{skyrmioistantoni}.

In this paper we construct another, so far unknown, example of a
composite BPS  soliton --- strings lying on domain walls.
These strings carry one half of the magnetic flux of the bulk strings
(hence, the name ``fractional"). From 3D perspective of the
wall world-volume theory they are akin to Polyakov's confining strings
of (1+2)-dimensional compact electrodynamics \cite{polyakov}.
A crucial element of our construction is the fact of the gauge field localization
on the wall observed in \cite{SY-abw} (a mechanism of such localization was
first discussed in \cite{DS}).

If we consider two distinct isolated supersymmetric vacua,
a domain wall configuration interpolating between them always exists.
Moreover,  sometimes there is more than one  interpolating configuration;
all of them must have one and the same tension, as the walls
under consideration are BPS-saturated.
Such a situation  can happen if the wall has a continuous
modulus or moduli  (in addition to the
translational modulus), which can be promoted to  field(s)   living  on the
wall world-volume, see e.g. \cite{SY-abw}.

Another possibility is that just a finite number of domain
walls exist interpolating between the two given vacua.
The first example of this type, two distinct walls, was found in SU(2) SQCD
\cite{sm}. In \cite{vafa} it was argued that
in SU($N$)  super-Yang--Mills with $\mathcal{N}=1$
the number of distinct ``minimal" domain walls
between vacua in any given pair of ``adjacent" vacua is $N$.
A weak coupling analysis  can be carried out by passing to a weakly coupled
Higgs phase through addition of fundamental matter
(see Refs. \cite{ritz1,ritz2}).

\begin{figure}[h]
 \centerline{\includegraphics[width=3in]{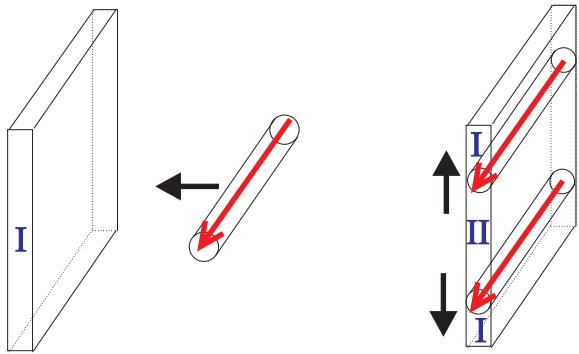}}
 \caption{\footnotesize The fate of the bulk flux tube parallel to a nearby
 domain wall of type I. In the bulk the string
 tension is of order $\xi$ while on the wall of order $\xi \beta/m  $.
 So for small $\beta$ the vortex is attracted by the wall.
 On the wall world-volume the string is a composite Sine-Gordon
 double kink; two Sine-Gordon kinks repel each other and so
 a region of type II domain wall will appear in between.
 At $\beta=0$ the kink disappears and the vortex just
 annihilates due to the fact that magnetic charges are in Coulomb
 phase on the wall world-volume.  }
\label{azzurro}
\end{figure}

Given the existence of a finite number of non-gauge-equivalent
walls connecting two given bulk vacua, it is natural at this point to address the problem of domain lines, which are   junctions between
two of the above walls. In this paper this problem is addressed
in the controllable regime of weak coupling. From the bulk perspective
the domain lines to be analyzed below
can be viewed as bulk strings put on the brane, see Fig.~\ref{azzurro}.
Since these domain lines carry magnetic flux they are
of an explicitly different class than those discussed in \cite{ritz2}.

Where does the magnetic flux come from?
In SQED {\em per se} one can introduce magnetic monopoles
\`a la Dirac, as probe external objects. Alternatively,
we could embed   the U(1) theory in the ultraviolet in an SU(2)
gauge theory, with SU(2) being spontaneously broken down to
U(1) at some high scale. Then the magnetic charges would appear as the
't Hooft--Polyakov  monopoles. Since the electric charges are condensed in both vacua,
to the left and to the right of the wall, a magnetic charge
placed in the bulk would form  a flux tube,
which would go to the wall in the perpendicular direction.
On the wall world-volume this tube splits in two distinct domain lines.
Each of them carries a half of the monopole magnetic  flux.
Since we have two {\em distinct} domain lines, it is possible to build  a static configuration where the monopole is represented by a junction of the two domain lines. This is
conceptually similar
to the confined monopoles of Refs. \cite{SY-mon,HT2,012,tong-mon,n=1*}.
Thus, the monopoles can be trapped not only by non-Abelian strings --- they can be nested inside
the walls too!

Our basic bulk model is  $\mathcal{N}=2$ SQED
with $N_f=2$ flavors and the Fayet--Iliopoulos (FI)
parameter $\xi$. This is the model used in \cite{SY-abw},
where the reader can find all relevant details.
In $\mathcal{N}=2$ SQED there are two equivalent
ways of introducing the FI parameter:
through the $D$-term and through superpotential \cite{HSZ,VY}.
We will introduce $\xi$ in the superpotential, along with a
small mass parameter $\beta$   breaking $\mathcal{N}=2$ down to $\mathcal{N}=1$.
We will work in the limit
\beq
\xi\gg m_A ,\,\,\beta\,,\qquad A=1,2\,,
\eeq
where $m_A$ are the mass parameters of the two matter hypermultiplets. An effective Lagrangian
emerging at low energies is that of the
sigma model with the Eguchi--Hanson manifold as the target space.
The  domain walls at $\beta=0$  in this problem
were studied previously in  \cite{tsm,jsm}.

The theory at $\beta=0$ has a moduli space of BPS walls
which can be parameterized by  a phase $\sigma$,
$$0<\sigma<2 \pi\,.$$
This is due to the fact that the symmetry of the
bulk  theory is U(1)$\times$U(1).
In each vacuum the local U(1) gets Higgsed while the other one remains unbroken.
However, it is spontaneously broken on the wall \cite{SY-abw}.
At $\beta=0$
the (probe) magnetic charges, monopoles, are confined in the bulk
since their magnetic flux gets squeezed into the
 Abrikosov--Nielsen--Olesen (ANO) flux tubes   \cite{ANO}.
 Inside the wall the charged matter condensates vanish,
 and the magnetic flux can spread all over the brane, so that the
 probe magnetic charges  on the wall world-volume
are in the Coulomb phase. This is the mechanism responsible for localization of
a gauge field on the domain wall.

Indeed, the scalar $\sigma$ can be dualized \`a la Polyakov
\cite{polyakov} into a gauge vector on
the wall world-volume.  The field strength tensor is given by (see e.g. Ref.~\cite{SY-abw}):
\beq
 F^{(2+1)}_{nm}=\frac{e^2_{2+1}}{2 \pi}\,
\epsilon_{nmk} \, \partial^k \sigma\,.
\label{eq2}
\eeq
The end-point of the ANO string on the wall
becomes a dual electric charge, the source of the
dual electric field in Eq. (\ref{eq2}) which, in the original description, was the magnetic field.

Now, what happens if we deform the above
theory by switching on $\beta\neq 0\,$?

Once we introduce a small $\beta$ perturbation
(we will consistently work in the first nontrivial order in $\beta$), the main
impact on the bulk theory is an explicit breaking of the global U(1).
As a result, the modulus $\sigma$ ceases to be a massless field.
A potential of the form
\beq
V=\frac{\beta ^2 \xi}{ m } \, \cos^2  \sigma +\mathcal{O}(\beta^3)
\label{eq3}
\eeq
develops.
The interaction (\ref{eq3})
is nonlocal in terms of the dual vector field (\ref{eq2}).

The potential (\ref{eq3}) lifts the moduli space
leaving us with two isolated vacua, at $\sigma =\pi/2$ and  $\sigma =3\pi/2$.
Thus, we have two degenerate
BPS domain walls to be referred to as walls of  type I and type II.
 The domain line is a ``wall on a wall", it divides
 the the wall into two domains -- one in which we have the type I wall from another
of type II wall. In terms of the effective world-volume description
the domain line is a sine-Gordon
kink in the effective world-volume theory (\ref{eq3}).
The sine-Gordon potential we obtain at $\beta\neq 0$ is in one-to-one correspondence
with Polyakov's sine-Gordon potential \cite{polyakov}.
There are three nuances, though: (i) Polyakov's  potential
is generated by instantons of the compact 3D electrodynamics;
(ii) in our case we deal with $\mathcal{N}=1$  sine-Gordon,
while  Polyakov's model was not supersymmetric;
(iii) the flux tube in Polyakov's model was unique, while we have two distinct types,
see below.

With the potential  (\ref{eq3}) generated, the Coulomb-like picture
of the dual field flux lines disappears, since the flux  gets collimated
inside the domain lines.
Note that in three dimensions there exist two distinct ways
in which long-range gauge potential can acquire a mass gap.
First, through a Chern--Simons term
\beq
S_{\rm CS}= \frac{1}{2 \pi} \, \epsilon_{nmk} \, A_n \partial_m A_k \,.
\eeq
as e.g. in Refs. \cite{vafa} and \cite{Tbrane,SYdual}.
 The Chern-Simon term is nonlocal
when written in term of the scalar filed $\sigma$.
Second, through a
sine-Gordon-type potential for $\sigma$ which, in turn,  is nonlocal in terms of the
dual vector potential.

The organization of the paper is as follows.
In Sect. 2 we
outline our bulk model and  discuss two vacua of the theory.
Sect. 3 is devoted to the  the sigma-model limit,
applicable at $\xi\gg m,\beta$. Here we   follow
the  parameterization introduced in Ref. \cite{jsm}.
Many   computations  carried out below   are simpler
in the sigma-model formalism.
In Sect. 4.1 the   domain walls at $\beta=0$ are reviewed, while in
in Sect. 4.2 we address the problem at $\beta\neq 0$.
Exact BPS solutions are found in the sigma-model limit
for  two distinct walls.
In Sect. 5 we derive the effective potential for the
quasi-modulus $\sigma$ assuming that $\beta$
is small. The potential is found in the leading
nontrivial order in $\beta$, namely $\beta^2$. In Sect. 6
we discuss the domain lines (2-wall
junctions in the nomenclature of Ref.~\cite{ritz2}). They are obtained
 as   kinks of the effective world-volume description. We show that
 there are two types of the domain lines, which thus have a junction of their own.
Sect. 7 contains a short summary and conclusions.

\section{Theoretical set-up}

\subsection{Lagrangian of $\mathcal{N}=2$ SQED with two flavors}

The bulk theory which we start from has the gauge group U(1)$_G$,
extended $\mathcal{N}=2$ supersymmetry,
and $N_f=2$ hypermultiplets of matter $\{Q^A,\,\,\tilde Q_A\},\,\,\, A=1,2$.
We endow it with the
Fayet--Iliopoulos parameter $\xi$ in the
superpotential. The latter contains the coupling ${\cal A} \sum_B \left( Q^B\,
\tilde Q_B\right)$ where ${\cal A}$ is the $\mathcal{N}=1$ chiral superfield,
the $\mathcal{N}=2$ superpartner of the  photon. It also contains the mass term
\beq
\tilde Q_A\, {\cal M}^A_B\, Q^B
\label{eq5}
\eeq
where $ {\cal M}$ is a mass matrix. If the matrix $\mathcal{M}_{ij}$ was Hermitean we could always diagonalize
it by unitary transformations of  $ Q^A$ and $\tilde Q_A$,
to obtain  a real diagonal matrix. In our case ${\cal M}^A_B$
is {\em not} Hermitean. Without loss of generality
we can choose it as follows:
\beq
{\cal M} =\left(\begin{array}{c|c}m & -\beta /\sqrt 2 \\\hline \beta  /\sqrt 2&-m \end{array}\right)\,,
\label{eq6}
 \eeq
 where $m$ and $\beta$ are real and positive parameters.
 If $\beta =0$ the superpotential (\ref{eq5})
 is compatible with  $\mathcal{N}=2$.
If $\beta\neq 0$ we  break $\mathcal{N}=2$ supersymmetry
down to  $\mathcal{N}=1$ once the Fayet--Iliopoulos term is
added to the theory.
 As was mentioned, we assume that $\beta \ll m$.

With these assumptions the  superpotential takes the form
\beq
\mathcal W=\frac{1}{\sqrt{2}} \, \mathcal A \, Q^B \tilde{Q}_B +
 m\left( Q^1 \tilde{Q}_1 -  Q^2 \tilde{Q}_2\right)
 +
 \frac{\beta}{\sqrt{2}} \left(Q^1 \tilde{Q}_2 - \tilde{Q}_1 Q^2\right)
-\frac{1}{2 \sqrt{2}}\,  \xi \,\mathcal A \,.
\label{eq7}
\eeq
The bosonic part of the action can be written as
\beq
 S=\int d^4 x \left\{ \frac{1}{4 g^2} F_{\mu \nu}^2 +
\frac{1}{g^2} |\partial_\mu a|^2 +\bar{\nabla}_\mu \bar{q}_A \nabla_\mu q^A +
\bar{\nabla}_\mu \tilde{q}_A \nabla_\mu \bar{\tilde{q}}^A
 +
V_D+ V_F \right\},
\label{n2sqed}
\eeq
where
$$
\nabla_\mu=\partial_\mu-\frac{i}{2}A_\mu, \,\,\,
\bar{\nabla}_\mu=\partial_\mu+\frac{i}{2}A_\mu\,,
$$
and $a$, $q$ and $\tilde q$ are the lowest components of the
chiral superfields $\mathcal{A}$, $Q$ and $\tilde Q$, respectively.
Moreover, the potential is the sum of the $D$ and $F$ terms,
\beq
V_D=\frac{g^2}{8}\left(|q^B|^2-|\tilde{q}_B|^2\right)^2\,,
\label{eq9}
\eeq
and
\beqn
&& V_F=\frac{1}{2} \left |q^1(a+\sqrt{2}m)-\beta q^2\right|^2+
\frac{1}{2}\, \left|\tilde{q}_2 ( a-\sqrt{2}m)-\beta \tilde{q}_1 \right|^2
\nonumber\\[3mm]
 &&+
\frac{1}{2} \left|(a +\sqrt{2}m) \tilde{q}_1+\beta \tilde{q}_2\right|^2+
\frac{1}{2} \left|(a -\sqrt{2}m) q^2+\beta q^1\right|^2+
\frac{g^2}{2}\left|\tilde{q}_A q^A-\frac{\xi}{2}\right|^2\,.
\nonumber\\
\label{potgauge}
\eeqn

At $m= \beta=0$ the theory at hand has two global symmetries: the SU(2)$_F$
flavor symmetry\,\footnote{Strictly speaking, the flavor group is
U(2); however, its  U(1) subgroup is gauged.}
and the SU(2)$_R$ $R$-symmetry,
 which is a general feature of $\mathcal{N}=2$ theories.

We pause here to comment on the Fayet--Iliopoulos parameter.
One can introduce a
generalized Fayet--Iliopoulos term $\vec{\xi}$
 (for details see \cite{VY})
as a triplet under  SU(2)$_R$.
The triplet is formed from the real and imaginary parts of the appropriate
$F$-term coefficient in the superpotential (see the last term in Eq. (\ref{eq7}))
and   the $D$-term coef\-ficient which is real.
In this paper we take
\beq
\im(\xi_F)=\xi_D=0 \quad\mbox{ and}\quad  \re (\xi_F)=\xi\,.
\label{eq11}
\eeq
We always can bring 
the Fayet--Iliopoulos parameter to this form by 
an SU(2)$_R$ rotation. Therefore, 
our parameter $\xi$ is real; we also assume  it to be the largest
in the scale hierarchy, $\sqrt \xi\gg m/g$.
We call $U(1)_R$ the $SU(2)_R$ subgroup which leaves
the Fayet-Iliopolous term in the form of Eq. (\ref{eq11}).

The flavor group  SU(2)$_F$ is broken down
to U(1)$_F$ by the parameter $m\neq 0$; the parameter
$\beta\neq 0$ breaks the flavor symmetry completely.
Furthermore,
SU(2)$_R$ is not broken by $m$ since this mass term preserves $\mathcal{N}=2$.
At the same time, SU(2)$_R$
is broken by $\beta\neq 0$.  A ${Z}_2$ subgroup of U(1)$_R \times$U(1)$_F$
generated by a $\pi$ rotation in both U(1)'s
is left unbroken by $\beta$ (this nontrivial property is somewhat
hidden in the gauge theory; it becomes  more transparent
 in the sigma-model formalism).

\subsection{Two vacua}
\label{twovac}

It is convenient to introduce two parameters,
\beq
\Omega = \frac{1}{2} \sqrt{\xi+\frac{m \xi}{\sqrt{m^2-\beta^2/2}}}, \qquad
\omega =\frac{1}{2} \sqrt{-\xi+\frac{m \xi}{\sqrt{m^2-\beta^2/2}}}\,.
\label{eq12}
\eeq
Note that  $\omega=0$ and $\Omega=\sqrt{\frac{\xi}{2}}$ at $\beta=0$.

It is not difficult to see that for $ \beta<\sqrt{2} m $ the theory has two vacua.
The first vacuum  is
\beqn
&& a=-\sqrt{2 m^2 -\beta^2},
\nonumber\\[3mm]
 && q^1=\bar{\tilde{q}}_1= \Omega, \quad
 q^2=-\bar{\tilde{q}}_2 =\omega \,.
 \label{eq13}
\eeqn
The second vacuum  is
\beqn
&& a=\sqrt{2 m^2 -\beta^2},
\nonumber\\[3mm]
 && q^1=-\bar{\tilde{q}}_1 =\omega, \quad
 q^2=\bar{\tilde{q}}_2 = \Omega\,.
 \label{eq14}
\eeqn
In order to check that the potential vanishes in these vacua we have to use the following arithmetic identity:
\beq
\sqrt{a \pm \sqrt{b}}=\sqrt{\frac{a+\sqrt{a^2-b}}{2}} \pm
\sqrt{\frac{a-\sqrt{a^2-b}}{2}}\, .
\label{eq15}
\eeq
At $\beta =0$ Eqs.~(\ref{eq13}) and (\ref{eq14})
reduce to two vacua considered in \cite{SY-abw}.
We will be interested in the domain walls interpolating between them.

\section{Sigma model description ($\beta, m \ll g \sqrt{\xi}$)}
\label{sigmamod}

\subsection{Constraints and parameters}
\label{constr}

Under the above choice of parameters we
can integrate out the scalar field $a$ and the gauge field. The low-energy
effective description is given by a sigma model with the target space
on $q^A,\tilde{q}_A$ which is
 the Eguchi--Hanson
manifold defined by the constraints
\beq
 |q^1|^2+|q^2|^2=|\tilde{q}_1|^2+
|\tilde{q}_2|^2 \,,\qquad  \tilde{q}_A q^A=\frac{\xi}{2} \,,
\label{eq16}
\eeq
in addition,  one  has to factor out  the action of the U(1) gauge group.

A convenient way to parameterize these constraints is explained
by Sakai and collaborators in Ref. \cite{jsm}. Let us introduce
the following functions:
\beq
f(r)=\sqrt{-\frac{\xi}{2} + \sqrt{r^2+(\xi/2)^2}}\,, \qquad
 g(r)=\sqrt{\frac{\xi}{2} + \sqrt{r^2+(\xi/2)^2}}\,.
 \label{eq17}
 \eeq
The Eguchi--Hanson manifold can be parameterized by the radial coordinate $r\geq 0$
plus three angles $\theta,\,\,\psi$ and $\varphi$ with
\beq 0
\leq r < \infty\,,  \quad  0\leq \theta \leq \pi\,,  \quad
 0\leq \varphi \leq 2 \pi\,,  \quad  0\leq \psi \leq 2 \pi \,.
  \label{eq18}
 \eeq
In terms of these functions and angle variables
the explicit expression is
\beqn
q^1
&=&
\frac{e^{i \varphi/2}}{2} \left( e^{i \psi/2} g(r) \cos{\frac{\theta}{2}}
+e^{-i \psi/2} f(r) \sin{\frac{\theta}{2}} \right),
\nonumber\\[3mm]
q^2
&=&
\frac{e^{-i \varphi/2}}{2} \left(e^{i \psi/2} g(r) \sin{\frac{\theta}{2}}
 -e^{-i \psi/2} f(r) \cos{\frac{\theta}{2}} \right),
\nonumber\\[3mm]
\tilde{q}_1
&=&
\frac{e^{-i \varphi/2}}{2} \left(e^{-i \psi/2} g(r) \cos{\frac{\theta}{2}}
 -e^{i \psi/2} f(r) \sin{\frac{\theta}{2}} \right),
 \nonumber\\[3mm]
\tilde{q}_2
&=&
\frac{e^{i \varphi/2}}{2} \left(e^{-i \psi/2} g(r) \sin{\frac{\theta}{2}}
 +e^{i \psi/2} f(r) \cos{\frac{\theta}{2}} \right).
  \label{eq19}
\eeqn
The generic sections at $r>0$ are three-dimensional
submanifolds parameterized by the three angles $\theta,\,\,\varphi$
and $\psi$.
The section at $r=0$ is a two-dimensional sphere;
only  $\theta$ and $\varphi$ are the relevant coordinates here,
while the $\psi$-circle  shrinks to zero.
A compact expression for $r$
in term of the squark vacuum expectation values (VEVs) is
\beq
\sqrt{4 r^2+\xi^2} =|q^1|^2 +|q^2|^2 +|\tilde{q}_1|^2 +|\tilde{q}_2|^2 \,.
\label{rosa}
\eeq
The following formulas are   useful, because they readily imply
the expressions for the angles $\theta,\,\,\varphi$ and $\psi$
in term of gauge invariant squark bilinears,
\beqn
q^1 \tilde{q}_1
&=&
\frac{\xi+\sqrt{ 4 r^2+\xi^2} \cos \theta}{4}
-\frac{i r \sin \theta \sin \psi}{2}\,,  \nonumber\\[3mm]
q^1 \bar{q}_2
&=&
 \frac{e^{i \varphi}(\xi \sin \theta-
2 r \cos \theta \cos \psi - 2 i r \sin \psi)}{4}\, .
\label{karelias}
\eeqn

\subsection{Metric}
\label{metric}

With this parameterization, the metric is given by the kinetic part of the
SQED Lagrangian
\beq
\mathcal{L}_{\rm k}=-|\partial_\mu q|^2-|\partial_\mu \tilde{q}|^2
+\frac{i}{2} A_\mu \left(\bar{q} \stackrel{\leftrightarrow}{\partial_\mu} q -
\bar{\tilde{q}} \stackrel{\leftrightarrow}{\partial_\mu} \tilde{q}\right)-
\frac{1}{4} A_\mu A^\mu\left(q\bar{ q} +\tilde{q}\bar{\tilde{q}}\right),
\label{metrica}
\eeq
where
\beq
\bar{q} \stackrel{\leftrightarrow}{\partial_\mu} q=\bar{q} (\partial_\mu q)
- (\partial_\mu \bar{q}) q\,.
\label{eq23}
\eeq
Since $A_\mu$ enters with no derivatives, we can eliminate
this field by virtue of the clas\-sical equation of motion,
\beq
 A_\mu=\frac{i \left(\bar{q}  \stackrel{\leftrightarrow}{\partial_\mu} q  -
\bar{\tilde{q}} \stackrel{\leftrightarrow}{\partial_\mu} \tilde{q}\right)}
{\bar{q} q+\bar{\tilde{q}} \tilde  q}
\,.
\label{eq24}
\eeq
The metric of the low-energy sigma model is (see Ref. \cite{jsm})
 \beqn
 \mathcal{L}_{\rm k}
  &=&
\frac{1}{\sqrt{4 r^2 +\xi^2}} \left\{
-(\partial_\mu r)^2 - (\partial_\mu \theta)^2 \left(r^2+\left(\frac{\xi}{2}\right)^2\right) \right.
\label{metrata} \nonumber\\[3mm]
&-&
\left.
(\partial_\mu \varphi)^2 \left(r^2+\left(\frac{\xi}{2}\right)^2 \sin^2 \theta\right)-
 (\partial_\mu \psi)^2 \, r^2 -
  (\partial_\mu \varphi)\,  (\partial_\mu \psi) \, (2 r^2 \cos \theta)
\right\}.
\nonumber\\
\label{eq26pp}
\eeqn
 At $\theta=0$ and $\theta=\pi$ there are some singularities
in the coordinates (at $\theta=0$ the part of the metric
which is bilinear in $d\varphi, d \psi$
 is of the form $(d\varphi+d\psi)^2$);
 at $\theta=\pi$ it has the form $(d\varphi-d\psi)^2$).
Another coordinate singularity takes place at $r=0$,
where the coefficients of the bilinears in $d \psi$  vanish.
All these singularities are only due to the coordinate
choice; the vacuum manifold is smooth everywhere.

\subsection{Potential}
\label{potential}

Now, let us integrate out the gauge field and the scalar field $a$.
We observe that the
 expression for $a$ gets a correction due to $\beta$,
 \beq a
 =
 \frac{\sqrt{2}m \left(|q^2|^2 +|\tilde{q}_2|^2-
|q^1|^2 -|\tilde{q}_1|^2\right)
 + \beta
 \left(\bar{q_1}  q^2- q^1 \bar{q_2}+
 \tilde{q}_1 \bar{\tilde{q}}^2 - \bar{\tilde{q}}^1
 \tilde{q}_2\right)}
{\sum \left(|q^A|^2 +|\tilde{q}_A|^2\right)}\,.
\label{eq27}
\eeq
Eliminating $A_\mu$ and $a$ we get the scalar potential in the form
\beqn
 V
 &=&
\frac{m^2(4 r^2 +  \xi^2 \sin^2 \theta )}{\sqrt{4 r^2 + \xi^2}}+
2 \sqrt{2} m\,  \beta\, r \left(\cos \theta \cos \varphi \cos \psi-
\sin \varphi \sin \psi\right)
\nonumber \\[3mm]
&+&\frac{\beta^2 \left(4 r^2+ 3 \left(\frac{\xi}{2}\right)^2 +2 \xi^2 \cos 2 \theta\left(\sin \varphi\right)^2+
\xi^2 \cos 2 \varphi\right)}{2 \sqrt{4 r^2 + \xi^2}}\,.
\label{potsigma}
\eeqn
This potential has two vacua, one at $\theta=0$ and $\varphi+\psi=\pi$,
the other at $\theta=\pi$ and at $\varphi-\psi=0$ (remember that at $\theta=0,\pi$
it is only a combination of $\varphi,\psi$ that is the coordinate).
Both of the vacua are at the same value of $r_0$,
\beq
 r_0=\frac{\beta \xi}{2 \sqrt{2 m^2-\beta^2}}\,.
 \label{eq29}
 \eeq
These are just the vacua discussed in Sect.~\ref{twovac}
in the sigma-model formalism.

The  U(1)  rotation of the phase $\varphi$ corresponds
to the  U(1)$_F$ group of the bulk gauge theory while the rotation
of the phase $\psi$ corresponds to U(1)$_R$.
It easy to verify that the part of the potential proportional to $m$
breaks SU(2)$_F\times$U(1)$_R$ down to U(1)$_F\times$U(1)$_R$.
At $\beta=0$ the potential (\ref{potsigma})
  does not depend on $\varphi$ and $\psi$. On
the other hand, the part which depends on $\beta$ breaks both   U(1)
factors. However, a ${Z}_2$ subgroup generated by a $\pi$
rotation in both  U(1)'s is left unbroken   by $\beta\neq 0$.

\section{BPS domain walls}
\label{bpsdw}

\subsection{$\beta=0$}
\label{betazero}

First of all
let us consider the case
$\beta=0,\,\, m \neq 0$.
This problem has been studied in detail in Refs. \cite{SY-abw,tsm,jsm},
therefore we only briefly  review it here.

The ansatz
$$
 q^A=\overline{\tilde{q}_A}\equiv \bar{\tilde{q}}^A
 $$
can be used in this case.
The BPS equations for the wall are
\beqn
&& \nabla_3 q^1 \pm \frac{1}{\sqrt{2}}\left(q^1
 (\bar{a}+\sqrt{2} m_1)\right)=0\,, \nonumber\\[3mm]
&& \nabla_3 q^2 \pm \frac{1}{\sqrt{2}} q^2
 \left(\bar{a}+\sqrt{2}m_2\right)
 =0\,, \nonumber\\[3mm]
&& \frac{1}{g}\partial_3 a \pm \frac{g}{\sqrt{2}}
\left(\bar{q}_B \,\bar{\tilde{q}}^B-\frac{\xi }{2 } \right)
=0\,.
\label{eq30}
\eeqn
The tension of the wall $ T_{\rm wall}$ is given by the central charge,
\beq
T_{\rm wall}=2\left| \mathcal{W}({\infty})-\mathcal{W}(-\infty)\right| =2 \xi m\, .
\label{eq31}
\eeq
This domain wall is a $1/2$ BPS solution of the
Bogomol'nyi equations \cite{Bo}: four of the eight supersymmetry generators
of the $\mathcal{N}=2$ bulk theory are broken by the soliton.

The bosonic moduli space is described by two collective coordinates.
One of them is of course associated with the translations in the $z$ direction. The other one is a compact  U(1) phase
parameter $0<\sigma<2 \pi$.  Indeed, if
$$
\{ q_1(z),\,\,q_2(z),\,\, a(z)\}
 $$
 is a solution, we can easily build another  solution which is {\em not}
gauge equivalent, namely
$$
\{  e^{i \sigma} q_1(z),\,\, q_2(z),\,\, a(z)\}\,.
 $$
This is due to a specific feature \cite{SY-abw}
of the breaking of the U(1)$_G\times$U(1)$_F$
 symmetry. In both   vacua only one squark
develops a VEV, and only one of the two  U(1)'s is broken.
 In each   vacuum the phase of the ``condensed" squark field is eaten by the Higgs mechanism.
 On the other hand, on the wall both the U(1)'s  are broken --- one phase is eaten by the Higgs mechanism, while the other becomes a Goldstone mode localized on the wall.

\begin{figure}[h]
\begin{center}
$\begin{array}{c@{\hspace{.2in}}c@{\hspace{.2in}}c} \epsfxsize=2.5in
\epsffile{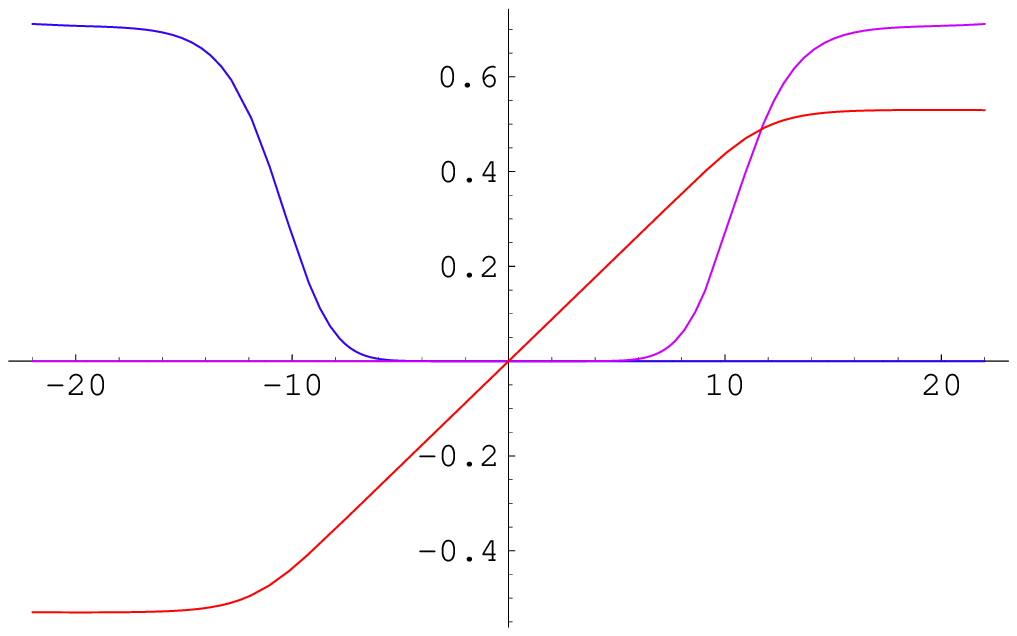}  &
     \epsfxsize=2.5in
    \epsffile{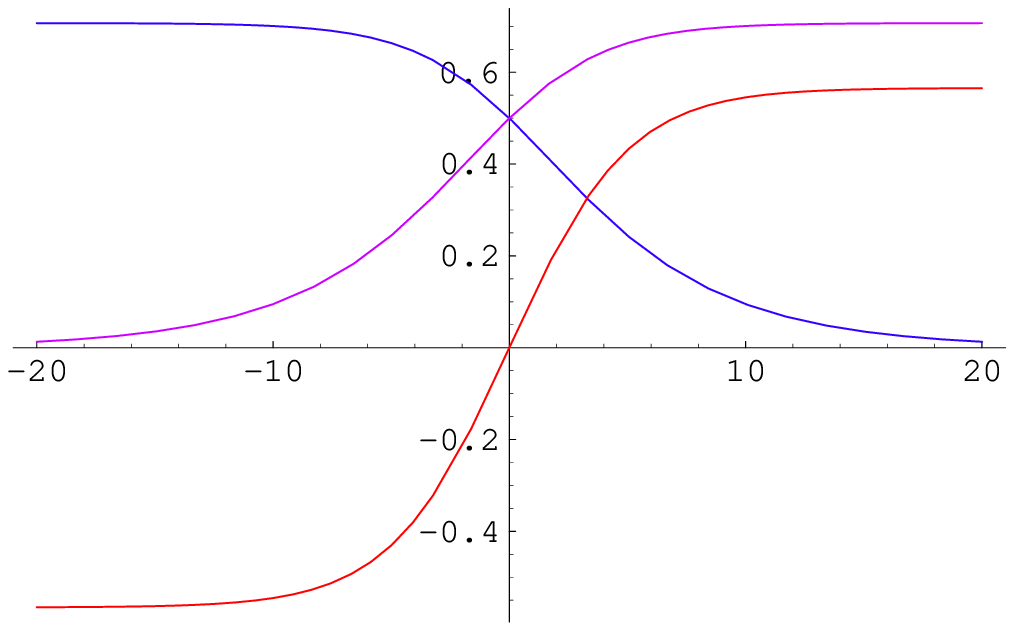}
\end{array}$
\end{center}
\caption{\footnotesize
The wall profile functions at $\beta=0$. the plot on the left represents
the large $m$ regime (a good approximation is
achieved already  at $m=3 \sqrt{\xi}$); on the right we see the small-$m$ sigma-model regime at $m=0.1 \sqrt{\xi}$. The field $q^1$ is depicted by  blue
color, $q^2$ by violet and $a$ by red. The units used for $a$ are different from those used for $q^1$.
}
\label{profiletti}
\end{figure}

If we change the relative ratio between $m$ and $\sqrt{\xi}$
the shape of the profiles functions changes.
In both  limits $m\gg \sqrt{\xi}$ and  $m\ll\sqrt{\xi}$ (the sigma-model
limit) a simple and compact solution can be found.
The $m\gg\sqrt{\xi}$ limit was studied in Ref. \cite{SY-abw}.
In this case the wall has a three-layer structure: there are two edges
of width $ \approx 1/\sqrt{\xi}$, where each of the squark
VEVs quickly drops to zero, and a large intermediate domain,
of width $2m/(g^2 \xi)$ (see Fig.~\ref{profiletti}, on the left).

The wall in the sigma-model limit was studied in Refs.~\cite{tsm,jsm}.
Let us summarize the results of these works, in our notation.
The potential is a function only of $r$ and $\theta$, namely,
\beq
V=\frac{m^2\left(8 r^2 + 2 \xi^2 \sin^2 \theta \right)}{\sqrt{4 r^2 + \xi^2}}
\,.
\label{eq32}
\eeq
The  two vacua are at $r=0, \,\, \theta=0$ and $r=0, \,\,\theta=\pi$.
Indeed, the wall also lives in the $S^2$ submanifold
at $r=0$. We can
use this {\em ansatz} (where
a profile function $\eta(z)$is introduced for   $\theta(z)$)
to obtain the following wall equations:
\beqn
q^1
&=&
\bar{\tilde{q}}^1=\sqrt{\frac{\xi}{2}} \left(\cos
\frac{\eta(z)}{2}\right)\,,
\nonumber\\[3mm]
q^2 &=&
\bar{
\tilde{q}}^2=\sqrt{\frac{\xi}{2}} \left(\sin \frac{\eta(z)}{2}\right)
e^{i \sigma}\,, \nonumber\\[3mm] a &=& m \sqrt{2}  \left(\sin^2
 \frac{\eta}{2} - \cos^2 \frac{\eta}{2}\right)= -m \sqrt{2} \, \cos
\eta\, .  \label{eq33}
\eeqn
The profile function $\eta(z)$ is
determined by   minimization  of the following energy functional:  
\beq
\int dz\, \mathcal{H}=\int dz\,
\left\{ \frac{\xi}{4} \left[(\partial_z \eta)^2 + \sin^2 \eta \,
(\partial_\mu \varphi)^2 \right]+  \xi m^2 \sin^2 \eta \right\}\,,
\label{eq34}
\eeq
which, in turn, implies the BPS equation
\beq
\eta'(z)= 2 m \sin ( \eta(z))\,,
\label{eq35}
\eeq
The solution of this equation is
\beq
\eta(z)=2 \arctan\, (\exp(2 m z))\,.
 \label{etta}
  \eeq
A compact expression can be written for the field profiles
 (see the plot   on the right-hand side of  Fig.~\ref{profiletti}),
\beqn
&& a=\sqrt{2} m \tanh ( 2 m z )\,,
\nonumber\\[3mm]
&& q^1=\bar{\tilde{q}}^1 = \frac{\sqrt{\xi}}{2}
\frac{e^{-mz}}{(\cosh 2 m z)^{1/2}} \,,
\nonumber\\[3mm]
 &&
  q^2=\bar{\tilde{q}}^2= \frac{\sqrt{\xi}}{2}
  \frac{e^{mz}}{(\cosh 2 m z )^{1/2}}\,.
  \label{eq37}
  \eeqn
Note that in the sigma-model limit ($m\ll \xi$ )
the wall thickness $R$ is of the order of $1/m$
(in the $m\gg \xi$ limit it was of the order of $m / (g \sqrt{\xi} )$).
This is interesting, because this means that $R$ is infinite
in both limits:  $m \rightarrow 0$ and   $m \rightarrow \infty$.
This happens because in both limits ($m\to 0$ and $\xi\to 0$)
certain states in the bulk theory become light.
 The minimal wall thickness  is at
$m=\mathcal{O}(\sqrt{\xi})$,
where we cannot use any of the above approximations.

It is possible to promote the moduli parameters to fields
depending on the world-volume coordinates.
The effective Lagrangian was found in Ref.
 \cite{SY-abw} using an explicit {\em ansatz} which takes into account
the world-volume dependence of $\sigma$. The bosonic part of
the world-volume action is
\beq
\int d^3 x \left\{ \frac{T_{\rm wall}}{2} \left(\partial_n z\right)^2
+\frac{\xi}{4 m} \left(\partial_n \sigma\right)^2\right\},
 \label{eq38}
\eeq
where the field $z$ describes the wall position in the perpendicular direction.
Furthermore, the
phase $\sigma$ can be dualized to a vector field in $(1+2)$ dimensions \cite{polyakov},
\beq
 F^{(2+1)}_{nm}=\frac{e^2_{2+1}}{2 \pi}\,
\epsilon_{nmk}\, \partial^k \sigma\,,
 \label{eq39}
\eeq
where
\beq e^2_{2+1}=\frac{2 \pi^2 \xi}{ m}\,.
 \label{eq40}
\eeq
Then the bosonic part of the world-volume action takes the form
\beq
\int d^3 x \left\{ \frac{T_{\rm wall}}{2}\, (\partial_n z)^2\,
-\frac{1}{4 e^2_{2+1}} (F_{mn}^{2+1})^2\right\}.
 \label{eq41}
\eeq
The full theory is $\mathcal{N}=2$ Abelian gauge theory in
three dimensions.

In both   vacua of the bulk theory the heavy trial
magnetic charges are  confined by
the Abrikosov--Nielsen--Olesen strings \cite{ANO}.
For example, in the vacuum where $a=-\sqrt{2} m$,
the {\em ansatz}
 $$
 q^2=\tilde{q}_2=0\,,\qquad \bar{\tilde{q}}^1=q^1
 $$
can be used to derive the following BPS
equations for the string:
\beq
\frac{1}{2} \,\epsilon_{ijk}\, F_{jk}-
\frac{g^2}{2} \left(|q^1|^2-\xi
\right)=0\,, \qquad \left(\nabla_1-
i \nabla_2\right) q^1=0\,.
 \label{eq42}
\eeq
The string tension is given by the $\{\frac{1}{2},\, \frac{1}{2}\}$
central charge \cite{GS}
\beq
T_{\rm string}=2 \pi \xi\, .
\eeq
The magnetic flux of the minimal-winding flux tube is $4 \pi$.
Absolutely similar equations can be written in the other vacuum,
$a=\sqrt{2} m$. The only difference is that here we now use the
{\em ansatz}
$$
q^1=\tilde{q}_1=0\,,\qquad q^2=\bar{\tilde{q}}^2\,,
$$
and a nontrivial equation is that  for the field $q^2$.
A BPS configuration exists in which the flux tube is perpendicular
to the wall and injects its magnetic flux into the wall
(this soliton is called boojum; it is studied in Refs.
\cite{SY-abw,j-boojums,rev1,ASY-boojums}).

So far, the magnetic field can freely propagate inside the wall.
String ends play the role of electric charges in the
effective U(1) gauge theory (\ref{eq41}) on the wall
\cite{SY-abw}. They are
 in the Coulomb phase on the wall  world-volume.
If we take a flux tube parallel to the wall, it will be attracted
to the wall in order to minimize the energy of the configuration.
If it reaches the wall it will dissolve, due to the fact that the
 charges are not confined on the wall.

 \subsection{Switching on $\beta \neq 0 $}
 \label{betanotz}

We know that if $\beta=0$ there is a moduli space
for the BPS walls parameterized by the
 phase $\sigma$,
 $$0\leq \sigma\leq 2 \pi\,.$$
Now let us generalize the BPS equations to take into account $\beta \neq 0$.
We expect to find just a finite number of BPS wall solutions,
because the U(1)$_G\times$U(1)$_F$  symmetry which
was responsible for the existence of the continuous modulus $\sigma$
is explicitly broken down to U(1) by the parameter $\beta$.
The  U(1) phase is eaten up by the
Higgs mechanism, and there is no extra global symmetry which can be used in order to build a Goldstone boson localized on the wall.

At $\beta \neq 0 $ the {\em ansatz} $q^A=\bar{\tilde{q}}^A$ is no longer
satisfied even in the vacua; therefore, it  is no more selfconsistent.
We have to keep
all the squark profiles as independent functional variables.
The Bogomolny completion (see Ref. \cite{Bo}) of the wall energy functional is
{ \small
\beqn
&&
\int d^3 x \, \mathcal{H} =
 \int dx_3 \left\{ \left|\nabla_3 q^1 \pm \frac{1}{\sqrt{2}}
 \left(\bar{\tilde{q}}^1\,
(\bar{a}+\sqrt{2}m_1)+ \beta \,\bar{\tilde{q}}^2 \right)
\right|^2\right.
\nonumber\\[3mm]
&&
 + \left|\nabla_3 \tilde{q}_1 \pm \frac{1}{\sqrt{2}}\bar{q}_1
 (\bar{a} +\sqrt{2}m_1) - \beta\,\bar{ q}_2 \right|^2
  + \left|\nabla_3 q^2 \pm \frac{1}{\sqrt{2}}\bar{\tilde{q}}^2
 (\bar{a}+\sqrt{2}m_2) - \beta \,\bar{\tilde{q}}^1\right|^2
 \nonumber\\[3mm]
&&   +\left|\nabla_3 \tilde{q}_2 \pm \frac{1}{\sqrt{2}}(\bar{q}_2
 (\bar{a}+\sqrt{2}m_2)+ \beta\,\bar{ q}_1)\right|^2
+ \left|\frac{1}{g}\partial_3 a \pm \frac{g}{\sqrt{2}}
\left(\bar{q}_B\,\,  \bar{ \tilde{q}}^B-\frac{\xi }{2 }
\right)\right|^2 \nonumber\\[3mm]
&&  \left. + \partial_3 \left[\left(\frac{a}{\sqrt{2}}  +
   m_B\right) \, q^B \tilde{q}_B-\frac{ \xi a}{2 \sqrt{2}}+
 \frac{\beta \, (q^1 \,\tilde{q}_2-\tilde{q}_1\, q^2)}{\sqrt{2}}
  \right]\mp {\rm c.c.}  \right\}.
 \label{stlouis2}
\eeqn }
The $D$-term constraint is
\beq
\nabla_3\left(|q|^2-|\tilde{q}|^2\right)^2 =0\,;
\eeq
it is recovered as a consequence of the   BPS
equations (\ref{stlouis2}).
The tension of the BPS wall is given by the appropriate central charge,
\beq
T_{\rm wall}
=2
\left|
\mathcal{W}({\infty})-\mathcal{W}(-\infty))\right|=2 \xi \sqrt{m^2-\beta^2/2}\,.
\label{tensio}
\eeq

 \begin{figure}[h]
\begin{center}
$\begin{array}{c@{\hspace{.2in}}c@{\hspace{.2in}}c} \epsfxsize=1.6in
\epsffile{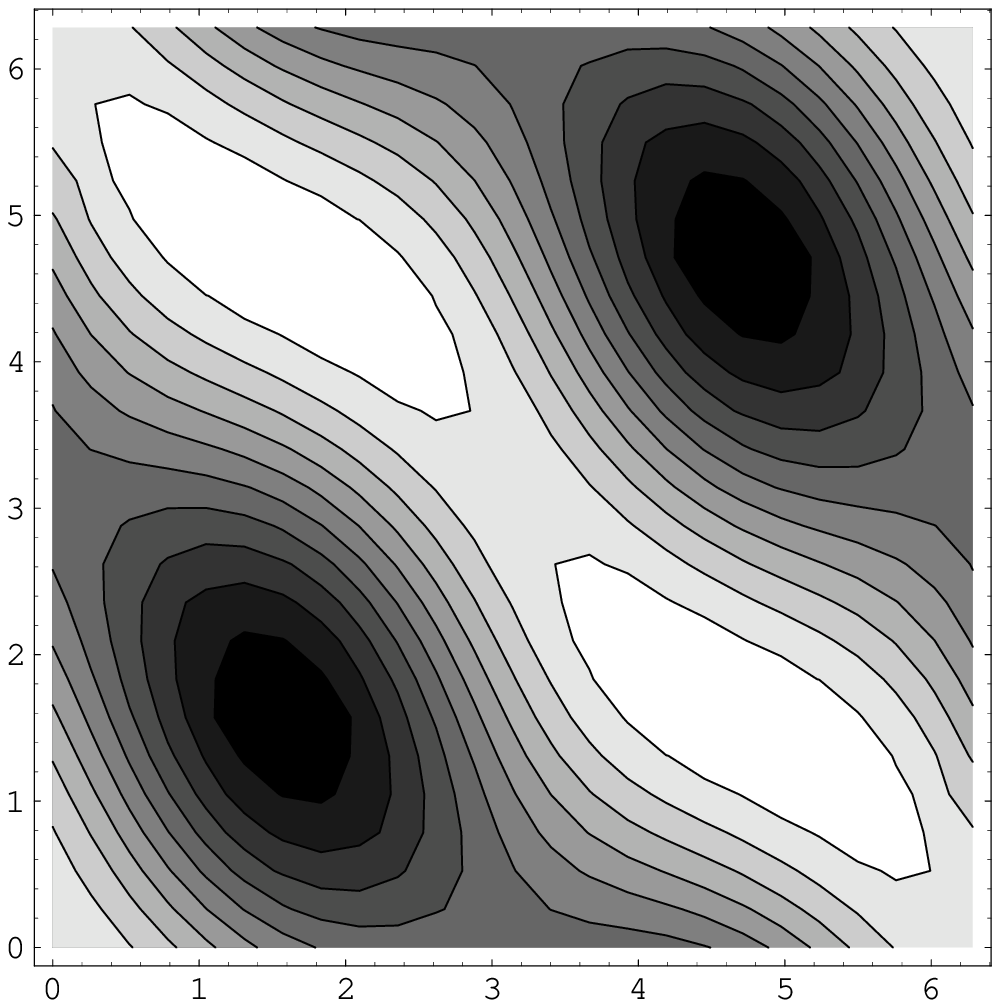} &
    \epsfxsize=1.6in
    \epsffile{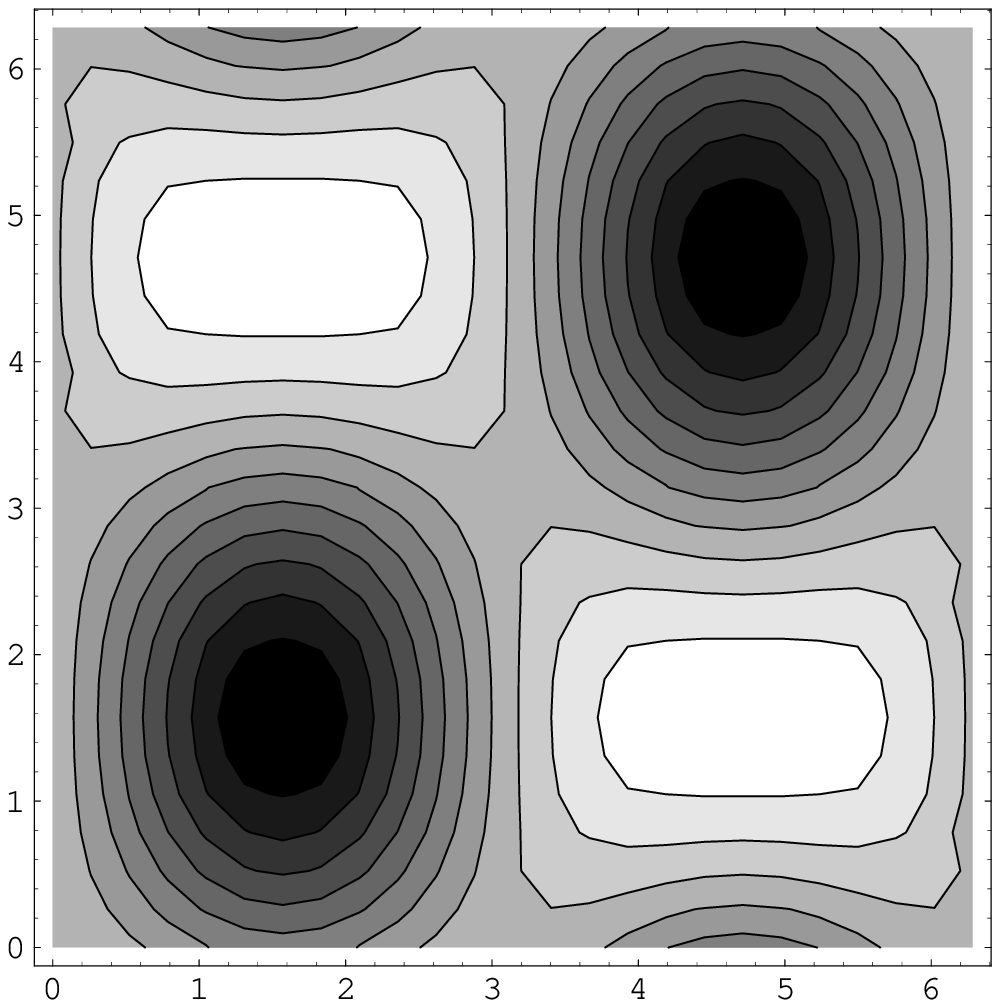} &
     \epsfxsize=1.6in
    \epsffile{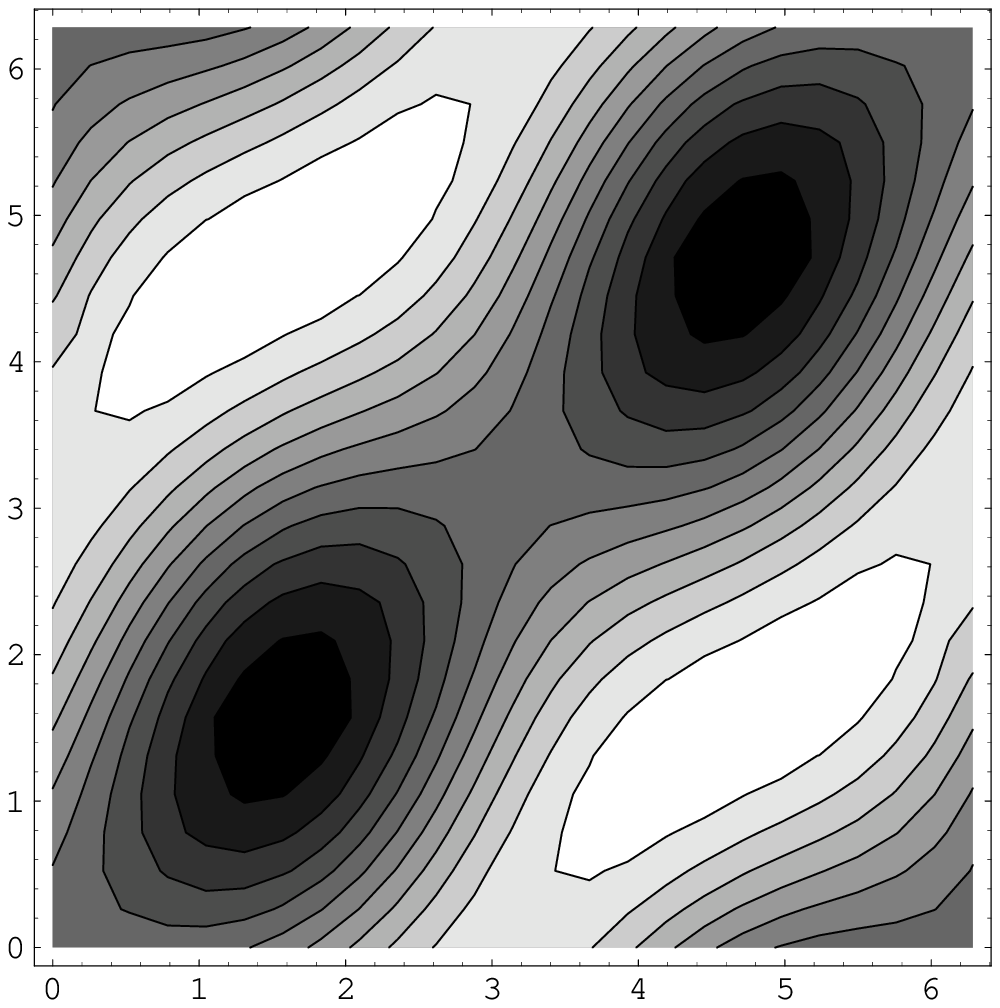}
\end{array}$
\end{center}
\caption{\footnotesize
Plot of the potential
at some sections at $r=r_0$ and at $\theta=\pi/3,\,\, \pi/2$ and $2 \pi/3$.
There are always two supersymmetric minima, one  at $\phi=\psi=\pi/2$
and the other at $\phi=\psi=3\pi /2 $. This is true
for all  values of $r$ and $\theta$, due to the ${Z}_2$
subgroup of U(1)$_F \times$U(1)$_R$ which is left unbroken by $\beta$.
}
\label{potenziale}
\end{figure}

In the gauge-theory approach the solution of the BPS equations are rather contrived.
It turns out that the problem is much easier in the sigma model
approach.  If we plot the potential in different slices at $r,\theta$ constant,
we find that there are always two minima, one at $\varphi=\psi=\pi/2$ and the
other at $\varphi=\psi=3 \pi/2$ (see Fig.~\ref{potenziale}).
Therefore,  this is the
appropriate {\em ansatz}  we have to use in order to find   two distinct wall solutions.
Two profile functions $r(z)$ and $\eta(z)$ are introduced
for the sigma-model coordinates $r$ and $\theta$.
Then the potential energy is
\beqn
  V(r,\eta)
  &=&
  \frac{1}{2 \sqrt{4 r^2+\xi^2}} \, \Big\{
(8 m^2 r^2-4 \sqrt{2}\, m \,\beta\, r  \,\sqrt{4 r^2+\xi^2}
\nonumber \\[3mm]
&+& \beta^2 \,(4 r^2+\xi^2)
+(2 m^2-\beta^2) \,\xi^2\, (\sin \eta)^2  \Big\}\,.
\label{eq47}
\eeqn
At $r=r_0$ the part of the potential which is constant in $\eta$ vanishes.
In terms of the profile functions the Bogomolny completion  of the wall energy is
\beqn
  \int d^3 x \, \mathcal{H}
  &=&
\int d^3 x \left\{
\left( \frac{\partial_z r}{\sqrt[4]{4 r^2+\xi^2}}-
\frac{2 m r - (\beta/\sqrt{2}) \sqrt{4 r^2+\xi^2}}{\sqrt[4]{4 r^2+\xi^2}}
\right)^2\right.
\nonumber\\[3mm]
&+&\left( \frac{\sqrt[4]{4 r^2+\xi^2}}{2}\, (\partial_z \eta) -
\frac{\sqrt{m^2-\beta^2/2}\,\, \xi \sin \eta }{\sqrt[4]{4 r^2+\xi^2}}
 \right)^2
\nonumber\\[3mm]
&-&\left. \xi \,\sqrt{m^2-\beta^2/2}
 \,\,( \partial_z\, \cos \eta )  \right\}\,.
 \label{bpssigma}
\eeqn
 The solution for $r(z)$ is just the constant value $r_0$. The solution
 for $\theta$ is completely similar to the case of vanishing $\beta$,
 \beqn
 \eta'(z)
 &=&
 (2m -\beta^2/m) \sin(\eta(z)) \,,
\nonumber\\[3mm]
 \eta(z)
 &=&
 2 \arctan\, \left[\exp\left((2 m-\beta^2/m) z\right)\right]\,.
\label{masa}
\eeqn
The tension is given by the total derivative term
\beq
T_{\rm wall} =2 \xi \sqrt{m^2-\beta^2/2}\,,
\eeq
which, of course, agrees with the result obtained from the central
charge of the gauge theory presented in Eq. (\ref{tensio}).

At this point we can return to  original SQED
and write the solutions for the squark fields.
For $\phi,\psi=\pi/2$
we have
\beqn
 q^1
&=&
 i \Omega \cos(\eta/2) + \omega \sin(\eta/2)\,,\quad
 q^2=i \omega \cos(\eta/2)+\Omega \sin(\eta/2) \,,
 \nonumber\\[3mm]
\tilde{q}_1
&=&
-i \Omega \cos(\eta/2) - \omega \sin(\eta/2)\,,\quad
 \tilde{q}_2=i \omega \cos(\eta/2)+\Omega \sin(\eta/2)
  \nonumber\\[3mm]
a
&=&
-\sqrt{2 m^2-\beta^2} \left(\cos \eta+i\frac{\beta}{\sqrt{2} m} \sin \eta\right)\,.
 \label{eq51}
 \eeqn
Moreover, for $\phi,\psi=3 \pi/2$ we have
\beqn
q^1
&=&
-i \Omega \cos(\eta/2) + \omega \sin(\eta/2)\,,\quad
 q^2=-i \omega \cos(\eta/2)+\Omega \sin(\eta/2) \,,
  \nonumber\\[3mm]
\tilde{q}_1
&=&
i \Omega \cos(\eta/2) - \omega \sin(\eta/2)\,,\quad
 \tilde{q}_2=-i \omega \cos(\eta/2)+\Omega \sin(\eta/2) \,,
   \nonumber\\[3mm]
a
&=&
-\sqrt{2 m^2-\beta^2} \left(\cos \eta-i\frac{\beta}{\sqrt{2} m} \sin \eta\right)\,.
 \label{eq52}
\eeqn
These solutions correspond to $\sigma=\frac{1}{2} \pi$ and $\sigma=\frac{3}{2} \pi$  of the $\beta=0$ case.
We checked by direct substitution
that they solve the gauge-theory BPS
equations (\ref{stlouis2})
in the limit in which the field $a$ is integrated out
(this is a cross-check for our calculations).
These wall solutions break spontaneously the ${Z}_2$
residual symmetry of the theory at $\beta \neq 0$.
This is the reason why we have two distinct walls.

\section{Unstable walls}
\label{unstable}

 When $\beta \neq 0$ there are just two stable domain wall solutions
 interpolating between the two vacua of the bulk theory.
 At generic $\sigma \neq \pi/2$ or $ 3\pi/2 $ we expect that
 the wall becomes unstable (perhaps, it is better to say, quasistable, at small $\beta$).
We will show that
from the standpoint of the world-volume theory this can be interpreted as an effective
potential with two degenerate minima
for the world-volume modulus $\sigma$.

Our purpose is to compute the tension of the unstable walls
in the first nontrivial order in $\beta$.
The nonconstant part of the wall tension will give
us the  effective potential of the world-volume effective theory.
To   this end we have to guess a field configuration
which interpolates between the two stable-wall configurations.
The following {\em ansatz} is a natural extension
of the  $\beta=0$ case, where  the exact answer is known:
\beqn
q^1
&=&
\frac{e^{i \sigma} \Omega \cos(\eta/2) + \omega \sin(\eta/2)}{A}\,,
\nonumber\\[3mm]
q^2
&=&
\frac{e^{i \sigma} \omega \cos(\eta/2)+\Omega \sin(\eta/2)}{A}\, ,
\nonumber\\[3mm]
\tilde{q}_1
&=&
\frac{e^{-i \sigma} \Omega \cos(\eta/2) - \omega \sin(\eta/2)}{  A}\,,
\nonumber\\[3mm]
\tilde{q}_2
&=&
\frac{-e^{-i \sigma}\omega \cos(\eta/2)+\Omega \sin(\eta/2)}{ A}\,.
\label{ansa}
\eeqn
As we will show below, the profile function $\eta(z)$  can  be
identified with the sigma-model coordinate $\theta$
only up to terms of order $\beta^2$.
The factor $A$ is introduced in order to maintain the sigma-model
constraint $|q^1|^2+|q^2|^2=|\tilde{q}_1|^2+|\tilde{q}_2|^2$.
This constraint implies
\beq
A=\left[ \frac{\sqrt{2}m+\beta \sin (\eta/2) \cos \sigma }{\sqrt{2}m-\beta \sin (\eta/2) \cos \sigma}\right]^{1/4}\,.
\label{eq54}
\eeq
As will be seen shortly, the above  {\em ansatz}  gives,
in the first nontrivial order
in $\beta$,  an elegant and compact result for the potential.

\begin{figure}[h]
\begin{center}
$\begin{array}{c@{\hspace{.2in}}c} \epsfxsize=2.4in
\epsffile{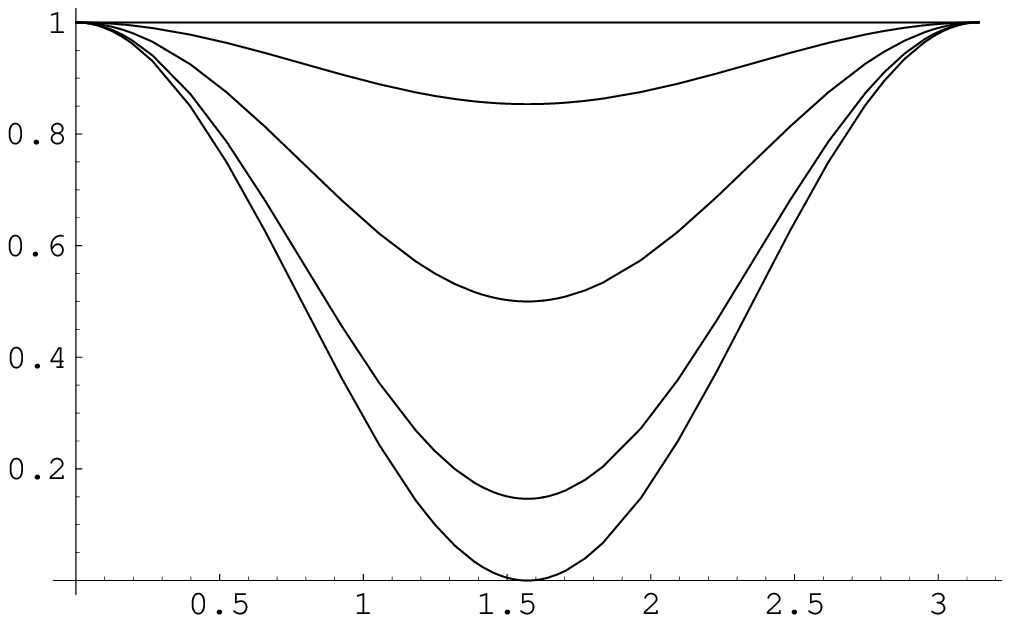}  &
     \epsfxsize=2.4in
    \epsffile{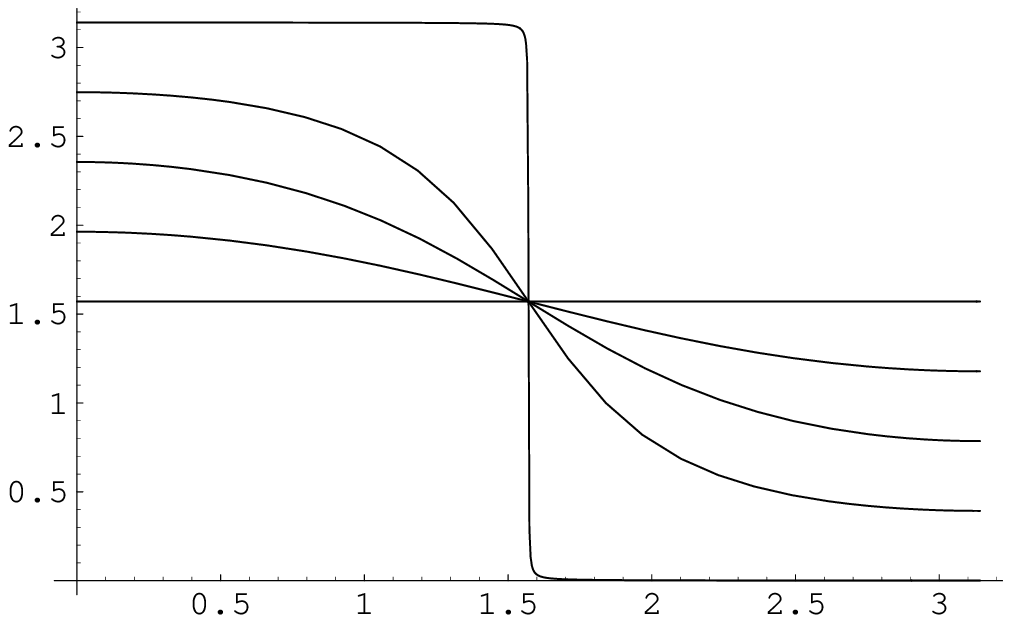}
\end{array}$
\end{center}
\caption{\footnotesize
Left: $r^2(\eta)$ for different values of $\sigma$.
  At $\sigma=\pi/2$ we have  $r$   constant.
At $\sigma=0$ and $\eta=\pi/2$ we have   $r=0$.
Right: $\psi(\eta)$ for different values of $\sigma$.
 At $\sigma=\pi/2$ we have   $\psi$  constant. At $\sigma=0$,
  $\psi$ tends to a step function, but this is not a problem
because $\sigma=0$  we have $r=0$.
}
\label{marpione}
\end{figure}

First of all, let us  translate Eqs. (\ref{ansa}) and (\ref{eq54})  in the sigma-model language.
This can be done using the expressions (\ref{rosa}) and (\ref{karelias}).
In the first relevant order in $\beta$ we have
\beqn
r
&=&
\frac{\xi \beta}{8 m } \sqrt{6+2 \cos (2 \eta) \cos (2 \sigma) +2 \cos(2 \eta) -2\cos(2 \sigma) }+\mathcal{O}(\beta^2)\,,
\nonumber\\[3mm]
\theta
&=&
\eta+\frac{\beta^2 \cos \eta \sin \eta \cos^2 \sigma}{4 m^2}+\mathcal{O}(\beta^3)\,,
\nonumber\\[3mm]
\phi
&=&
\sigma+\mathcal{O}(\beta^2)\,,
\nonumber\\[3mm]
\sin \psi
&=&
\frac{2 \sqrt{2} \sin \sigma}{\sqrt{6+2 \cos (2 \eta) \cos (2 \sigma) +
2 \cos(2 \eta) -2\cos(2 \sigma)}}+
\mathcal{O}(\beta)\,.
\label{barbagianni}
\eeqn
The plot of the functions $r(\eta)$ and $\psi(\eta)$
for some values of $\sigma$ is presented in Fig.~\ref{marpione}.

We will solve the problem for the effective potential for the modulus $\sigma$ at order $\beta^2$. To this end we have to compute the function $\eta(z)$ in
 the second order in $\beta$.
Let us write the energy density as a function of $\sigma$ and $\eta(z)$
in the second order in $\beta$.
The kinetic part  can be  easier  found   using
Eq. (\ref{barbagianni})
 in the sigma-model metric  (\ref{metrata}). The kinetic part takes the
form \beq
\left(\frac{\xi}{4}
+\beta^2\frac{10+6 \cos(2 \eta) +2 \cos(2 \sigma)+
6 \cos(2 \eta) \cos(2 \sigma)}{128 m^2}
\right)\left(\frac{d\eta}{dz}\right)^2\,.
\label{eq56}
\eeq
The potential part  can be easier found   in the original bulk gauge
theory, substituting   Eq. (\ref{ansa}) in Eq. (\ref{potgauge}).
The potential part is
\beq
(\sin(\eta))^2 \left(m^2 \xi
-\frac{\beta^2 \xi}{32}
 (10+6 \cos(2 \eta) -14  \cos(2 \sigma)+
6 \cos(2 \eta) \cos(2 \sigma))\right).
\label{eq57}
\eeq
Now, a second order equation can be obtained for the profile
function $\eta(z)$,
{\small
\beqn
&& \eta''\left(32 m^2+\beta^2(10+6 \cos(2 \eta) +2 \cos(2 \sigma)+
6 \cos(2 \eta) \cos(2 \sigma)) \right)
\nonumber
  \\[3mm]
&& +\,4 \, \xi m^2 (\sin 2 \eta) \left(
-16 m^2 +\beta^2(2+6 \cos(2 \eta) -10 \cos(2 \sigma)+
6 \cos(2 \eta) \cos(2 \sigma))
\right)
\nonumber\\[3mm]
&&\,+\,\eta'(24 \beta^2 (\cos \sigma)^2 \sin 2 \eta)=0\,.
\label{eq58}
\eeqn}
We will find the solution to the above equation iteratively in $\beta$.
Let us write
\beq
\eta=\eta_0 + \beta^2 \eta_2 + \mathcal{O}(\beta^3) \,,
\label{eq59}
\eeq
where from Eq.~(\ref{etta}) we have
\beq
\eta_0=2 \arctan\, (e^{2 m z})\,.
\label{eq60}
 \eeq
Substituting Eqs. (\ref{eq59}) and
(\ref{eq60}) in Eq. (\ref{eq58})
we arrive at a linear differential equation for $\eta_2$,
\beqn
&& 2 \left(\cosh (2 m z)\right)^4  \eta_2'' +
4 m^2 \left(\cosh(2 m z)\right)^2 \left(3-\cosh(4 m z )\right) \eta_2
\nonumber
  \\[3mm]
&&
+ 2 \sinh(2 m z) \left(-2+ 15 (\cos \sigma)^2-2 \cosh(4 m z)\right)=0\,,
\label{eq61}
\eeqn
which can be readily solved exactly
\beq
\eta_2 =-\frac{1}{m^2}\,\,
\frac{e^{2 m z}\left(15 \cos^2 \sigma +8(1+e^{4 m z}) m z\right)}
{4(1+e^{4mz})^2}\,.
\label{eq62}
\eeq
Note that for $\sigma=\pi/2,\,\, 3\pi/2 $ Eq.~(\ref{eq62}) reduces to
Eq. (\ref{masa}) in the second order in $\beta$.
The final  result for $\eta$ thus takes the form
\beq
\eta(z)=2 \arctan (e^{2 m z})-\frac{\beta^2}{m^2}\,
\frac{e^{2 m z}\left(15 \cos^2 \sigma +8(1+e^{4 m z}) m z\right)}
{4(1+e^{4mz})^2}+\mathcal{O}(\beta^3)\,.
\label{eq63}
\eeq

Now we can substitute
the solution (\ref{eq63}) back in the energy density  and
integrate in the $z$ direction. In this way we find
 the tension of
 the unstable wall at generic values $\sigma$, namely,
\beq
T_{\rm wall} (\sigma)=
2 \xi m + \frac{\beta ^2 \xi}{2 m }\, \cos (2 \sigma)+\mathcal{O}(\beta^3) \,.
\label{eq64}
\eeq
Of course, the minima at $\sigma=1/2 \pi ,\,\, 3\pi/2 $ correspond to the
two degenerate stable BPS walls.
The unstable-wall tension versus $\sigma$  reduces to
\beq
T_{\rm wall} (\sigma)=T_{\rm wall 0} +V(\sigma)+\mathcal{O}(\beta^3)\,,
\label{eq65}
\eeq
where
\beq
T_{\rm wall 0}=2 \xi m - \frac{\beta^2 \xi}{2 m}
\label{eq66}
 \eeq
is the tension of the stable walls
(note that this coincides with  Eq.~(\ref{tensio}) in  the first order in $\beta^2$)
and
\beq
V(\sigma)=\frac{\beta ^2 \xi}{m } \, (\cos  \sigma )^2
\label{eq67}
\eeq
is the world-volume potential for the field $\sigma$.
One can (and should) supersymmetrize the world-volume Lagrangian with
no difficulty.

\section{Domain line as a sine-Gordon kink}
\label{dline}

\begin{figure}[h]
\begin{center}
$\begin{array}{c@{\hspace{.2in}}c@{\hspace{.2in}}c} \epsfxsize=2.4in
\epsffile{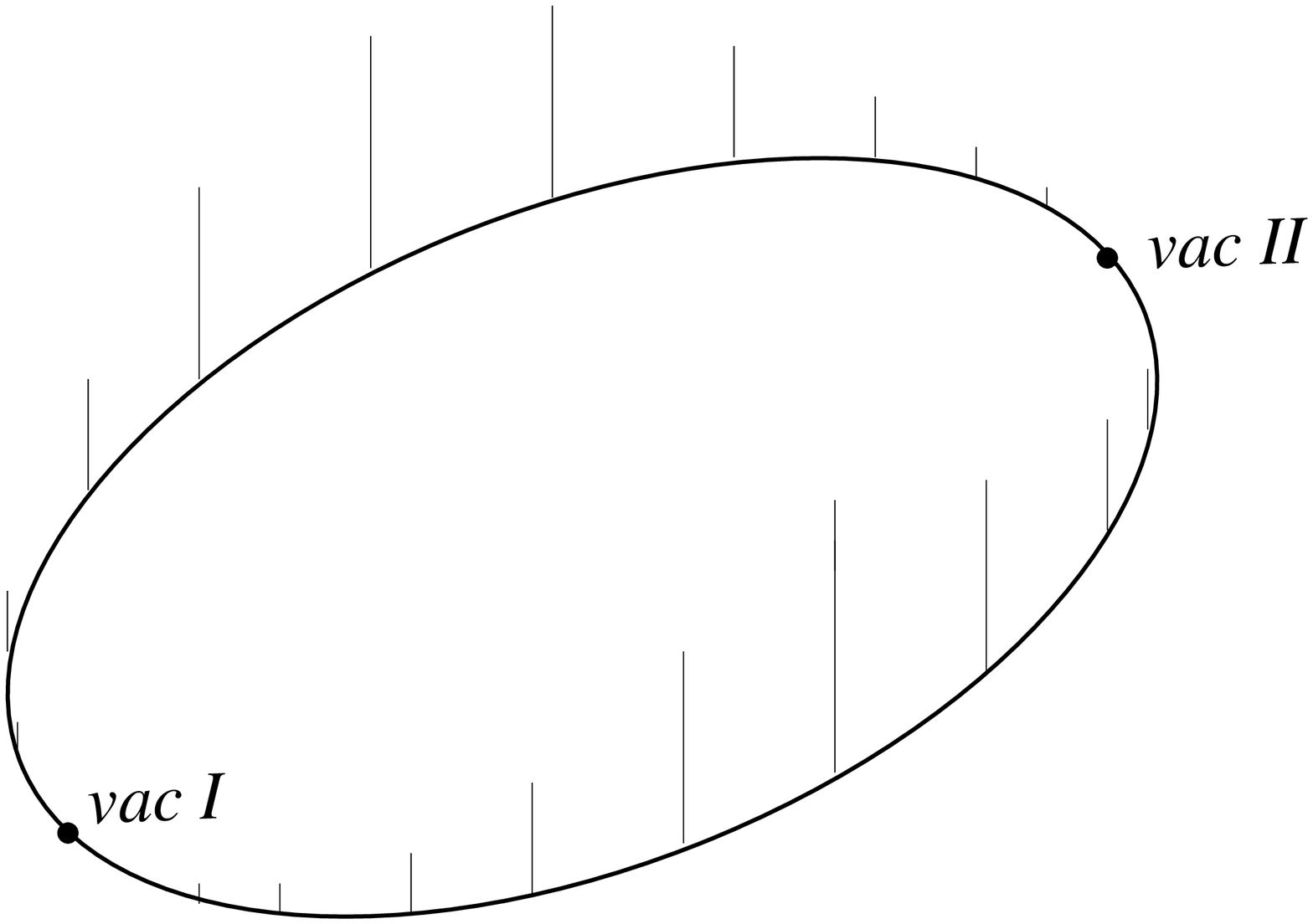}  &
     \epsfxsize=2.2in
    \epsffile{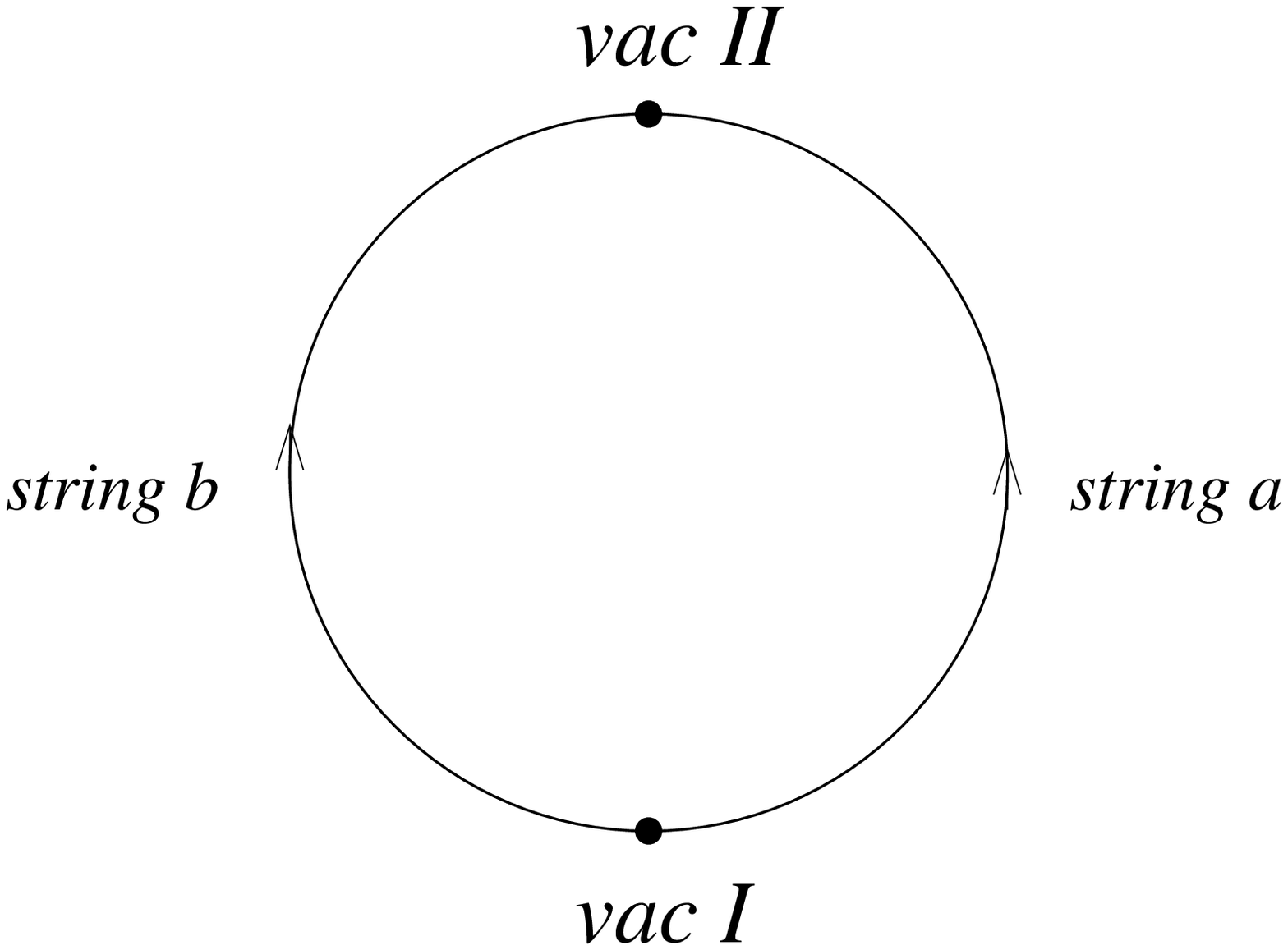}
\end{array}$
\end{center}
\caption{\footnotesize
 The potential energy for the field $\sigma$ defined on a circle
 $\sigma \in [0, 2\pi )$. The potential energy is presented by the height of the vertical
 lines. Two minima with the vanishing potential energy are vacua $I$ and $II$, respectively. There are two types of strings in the world-volume theory.}
\label{dmnlines}
\end{figure}

The action of the effective theory on the wall
world-volume,  in the first approximation in $\beta^2$, is that of
the supersymmetric sine-Gordon model,
with 2 supercharges. (We leave aside the translational modulus
and its superpartner.)
The bosonic part of the Lagrangian
(at $\mathcal{O}(1)$ in $\beta$ for the kinetic part and
 $\mathcal{O}(\beta^2)$ for the potential) is
\beq
S = \int d^3 x \left[ \frac{\xi}{4 m} (\partial_n \sigma)^2
- \frac{\beta^2 \xi}{ m} \cos^2 \sigma  \right].
\label{eq68}
\eeq
We can choose the vacuum at  $\sigma=- \pi/2$ as vacuum I and at
$\sigma=\pi/2$ as vacuum II (see Fig.~\ref{dmnlines}).

Now it is time to discuss the sine-Gordon kinks interpolating between these two vacua.
From Fig.~\ref{dmnlines} it is obvious that
there are two distinct kinks
which represent trajectories starting in   vacuum I and ending in vacuum II.

Let us denote the world-volume spatial coordinates as $\{x,\,\,y\}$.
The   domain line  is assumed to be oriented
in the ${y}$ direction. Therefore the field $\sigma$ in the kink solution
will depend on $x$.

The Bogomol'nyi completion of the energy functional can be represented as
\beqn
\mathcal{H}
&=&
\int d^2 x \left( \frac{\xi}{4 m} (\partial_n \sigma)^2
+ \frac{\beta^2 \xi}{ m} \cos^2 \sigma  \right)
\nonumber\\[3mm]
&=&
\int d^2 x \left\{ \frac{\xi}{2 m}
\left(   \frac{1}{\sqrt{2}} (\partial_x \sigma)
\mp \sqrt{2} \beta \cos \sigma  \right)^2 \pm \frac{ \xi \beta }{ m}
\partial (\sin \sigma) \right\}\,.
\label{eq69}
\eeqn
It is not difficult to obtain   two distinct kink
solutions  interpolating between vacuum I and vacuum II,
\beq
\sigma=2 \arctan\, (\exp(2 \beta x)) -\frac{\pi}{2}\,,
\label{eq70}
\eeq
(this solution has $\sigma(x \rightarrow -\infty)=-\pi/2$,
$\sigma(x \rightarrow +\infty)=\pi/2$, let us call it
$a$-string) and
\beq
\sigma=-2 \arctan\, (\exp(2 \beta x)) -\frac{\pi}{2}\,,
\label{eq71}
\eeq
(this solution has $\sigma(x \rightarrow -\infty)=-\pi/2$, 
$\sigma(x=+\infty) \rightarrow -3\pi/2 $, let us call
it $b$-string).
Of course, there are  two domain lines
interpolating between vacuum II and I --- antilines ---
which can be obtained by replacing  $x\rightarrow -x$. We denote them as
$\bar{a}$ and $\bar{b}$.

The transverse size of these objects is of the  order of $1/\beta$ and the tension is
\beq
T_{\rm dl}=\frac{2  \beta \xi}{m}\,.
\label{eq72}
 \eeq
Note that at this order in $\beta$
 the kink is BPS-saturated in the world-volume
theory.  We expect that the domain line saturates the $\{ \frac{1}{ 2},\,\, \frac{1}{ 2}\}$ central charge of the $(1+3)$-dimensional theory and that these solitons
remain  BPS-saturated to all orders in $\beta$.

  \begin{figure}[h]
 \centerline{\includegraphics[width=3in]{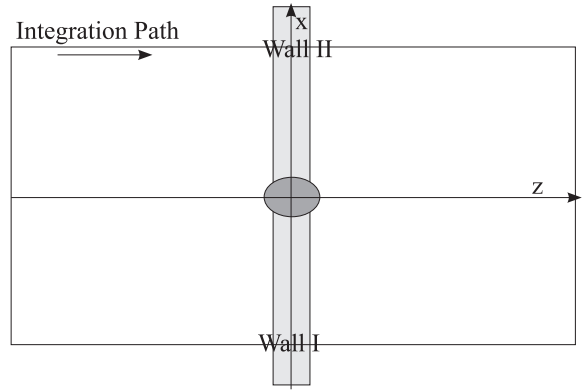}}
 \caption{\footnotesize Computation of the domain-line magnetic flux.}
\label{marziano}
\end{figure}

From the standpoint   of $(1+3)$-dimensional theory
 the above domain lines are  strings
 each of which carries the magnetic flux
 that is
 $1/2$ of the magnetic flux
of the flux tube living in the bulk.
It is easy to calculate the magnetic flux in the nonsingular gauge; it is given by
line integral over the vector potential.
Let us consider a rectangular path   depicted  in Fig.~\ref{marziano}.
On the left-hand side of the rectangle, at $z\rightarrow -\infty$, we choose
$q^A=e^{i \sigma(x)} q^{A}_{\rm vac}$ where $q^{A}_{\rm vac}$ are the squark VEVs.
On the right-hand side we choose to keep $q^A$ constant; on the two horizontal
sides we use the gauge in which $A_z=0$ (this can be done for each of the wall solutions).
 In order to keep the covariant derivatives vanishing on the left-hand side
 we take
  \beq
  A_x=2 \partial_x \sigma(x)\,,
   \eeq
where $\sigma(x\rightarrow -\infty)=-\pi/2,
\sigma(x\rightarrow +\infty)=\pi/2 $ for the $a$-string and
$\sigma(x\rightarrow -\infty)=-\pi/2,
\sigma(x\rightarrow +\infty)=-3 \pi/2 $ for the $b$-string.
On the other sides we have $A_k=0$.
The magnetic flux is then given by
\beq
\mathcal{F}_{\rm dl}=\oint \vec{A}\cdot d \vec{x}=2 ( \sigma_{x\rightarrow \infty}-
\sigma_{x\rightarrow -\infty} )=\pm2 \pi,
\eeq
where the $+$ sign is for the $a$-string and the $-$ sign is for the $b$-string.
On the other hand, if we compute in a similar way the flux of the bulk
Abrikosov--Nielsen--Olesen string, we find  $$ \mathcal{F}_{\rm ANO}=4 \pi\,.$$

The two different solutions we obtained --- the domain lines  $a$ and $b$
--- correspond to two possible orientations of the magnetic flux in the
${y}$ direction. An isolated string can not be taken out of the wall, because it connects two different vacua of the wall world-volume theory. On the other hand, if we take a bound state of an $a$-line and a $\bar{b}$-line, this configuration
interpolates, through one winding,
between one and the the same vacuum of the world-volume theory.
 Indeed, this bound state carries the same magnetic flux as
the flux tube living in the bulk, and, therefore,  can be pulled out from the wall.
If we take a bound state of an $a$- and   $\bar{a}$-lines we get a topologically trivial
configuration:  $a$ and   $\bar{a}$  domain lines  annihilate each other.

\begin{figure}[h]
 \centerline{\includegraphics[width=3in]{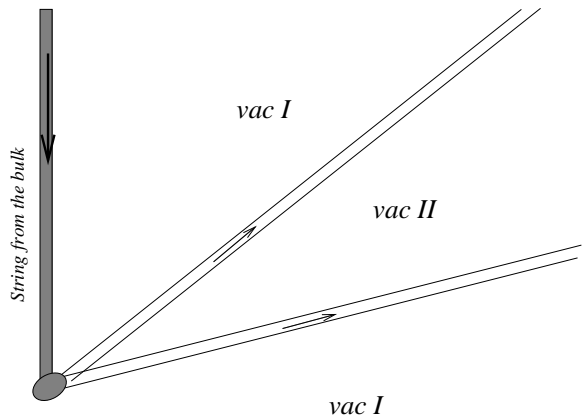}}
 \caption{\footnotesize Two type of strings in the world-volume theory.
 A bulk string is a bound state of the $a$ and the $\bar{b}$ domain lines.}
\label{asys3}
\end{figure}

In our U(1)  theory {\em per se} there are no magnetic monopoles,
but we could introduce these objects  by
embedding the  U(1)  theory in the ultraviolet in an SU(2)
gauge theory.  If we consider a probe magnetic monopole in the bulk,
 its magnetic flux will be squeezed in a bulk string which
 orients itself perpendicular to the domain wall.
 At a certain point this flux tube hits the wall (Fig.~\ref{asys3}).
This junction is seen in the world-volume as the
source of two domain lines, $a$ and $\bar b$, each carrying
one half of the flux of the bulk string.  A static stable configuration
appears when two domain lines point in the opposite directions on the
wall.

In the (2+1) dimensional effective theory on  the wall
the end-point of the bulk ANO string is seen as an electric charge
coupled to the (dual)  U(1) gauge field
at $\beta =0$
\cite{SY-abw,ASY-boojums,Tbrane,SYdual}. However, at non-vanishing $\beta$ we
can no longer dualize the phase $\sigma$  into the U(1) gauge field 
because the cosine term in   (\ref{eq68}) becomes
non-local. Therefore, at nonvanishing  $\beta$ it is better to use the
sine-Gordon formulation  (\ref{eq68}). In this formulation the
end-point of the bulk ANO string looks as a vortex of the field
$\sigma$ in the $(x,y)$-plane. At $\beta=0$ this vortex is given
by the solution \cite{SY-abw}
\beq
\sigma=\alpha,
\label{vortex}
\eeq
where $\alpha$ is the polar angle in the $(x,y)$-plane. At nonvanishing
$\beta$ the vortex $\sigma(\alpha)$ has two jumps --- each by the angle
$\pi$ --- tied up to the  directions of the $a$ and $\bar b$ domain lines.
We see that at $\beta\ne 0$ we  have no local description
of the junction of the bulk ANO string
with $a$ and $\bar b$ domain lines.

\begin{figure}[h]
\centerline{\includegraphics[width=3in]{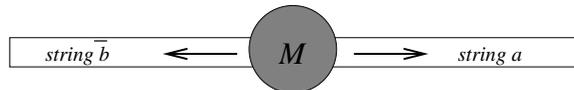}}
 \caption{\footnotesize A static configuration where a monopole
 is the junction of two domain lines is possible on the wall
 world-volume.}
\label{asys4}
\end{figure}

Now we can move the heavy trial monopole (the source of the bulk ANO
string) towards the wall. The length of the bulk string becomes
shorter and, eventually, when the monopole hits the wall, the bulk string
disappears. The configuration becomes a junction of the monopole
 trapped inside the wall with the
$a$ and $\bar b$ domain lines.
It is depicted, on the $(x,y)$-plane, in Fig.~\ref{asys4}.
This construction is analogous (conceptually rather than technically)
to non-Abelian  strings in the bulk, whose junction represents
a monopole trapped on the string \cite{tong-mon,SY-mon,HT2,012}. If a
pair of such wall-trapped monopoles (more precisely, an $M\bar M$ pair)
is nailed at two distant  fixed points on the 2D plane,
the strings stretched between them are slightly curved because of
their repulsion due to
their non-BPS nature.

\section{Conclusions}

In this paper we presented an explicit construction of the
domain line solitons in a weakly coupled $\mathcal{N}=2$ Abelian theory.
In our example the world-volume effective
theory describing  physics on the wall is the sine-Gordon model in $2+1$
dimensions, with two degenerate vacua. The domain line is a kink
of the sine-Gordon effective Lagrangian. Magnetic charges on the wall
are confined by these objects; the magnetic flux carried by them
is one half of the one carried by the the Abrikosov--Nielsen--Olesen
flux tubes in the bulk.

The effective description of the world-volume physics found in
this theory is different from the Chern-Simon description proposed for
the domain-wall world-volume
theory in  $\mathcal{N}=1$ super-Yang--Mills in Ref. \cite{vafa},
and we know why.
In particular, the Chern-Simon term is local only in terms of the dual
gauge field
description while, on the other hand, the sine-Gordon description is local only in
terms of the phase  field $\sigma$.

Also we would like to note that instantons inside domain wall in 
five dimensional gauge theory were recently discussed in [20].
In the effective world volume theory these instantons are seen as 
skyrmions. This has certain analogy with our results: string-like 
objects inside the  domain wall are seen as kinks in the effective 
sine-Gordon theory on the wall.

\section*{Acknowledgments}

We are grateful to Vassilis Spanos and  Joel Giedt for useful discussions.
The work  of R. A. and M. S. is
supported in part by DOE grant DE-FG02-94ER408.
The work of A. Y. is supported by Theoretical Physics Institute
at the University of Minnesota and also by INTAS grant
No. 05-1000008-7865 and RFBR grant No. 06-02-16364a.


\begin{thebibliography}{100}

\bibitem{SW1}
N.~Seiberg and E.~Witten,
Nucl. Phys. {\bf B426}, 19 (1994),
(E) {\bf B430},  485 (1994) [hep-th/9407087].

\bibitem{SW2}
N.~Seiberg and E.~Witten,
Nucl. Phys. {\bf B431}, 484  (1994)
[hep-th/9408099].

\bibitem{HT}
 A.~Hanany and D.~Tong,
  JHEP {\bf 0307}, 037 (2003)
  [hep-th/0306150].

\bibitem{ABEKY}
  R.~Auzzi, S.~Bolognesi, J.~Evslin, K.~Konishi and A.~Yung,
  Nucl.\ Phys.\ B {\bf 673} (2003) 187
  [hep-th/0307287].

 \bibitem{junct}
  G.~W.~Gibbons and P.~K.~Townsend,
  Phys.\ Rev.\ Lett.\  {\bf 83} (1999) 1727
  [hep-th/9905196];
   S.~M.~Carroll, S.~Hellerman and M.~Trodden,
  Phys.\ Rev.\ D {\bf 61} (2000) 065001
  [hep-th/9905217].

 \bibitem{GS}
  A.~Gorsky and M.~A.~Shifman,
  Phys.\ Rev.\ D {\bf 61}, 085001 (2000)
  [hep-th/9909015].

\bibitem{StV}
  M.~A.~Shifman and T.~ter Veldhuis,
  Phys.\ Rev.\ D {\bf 62} (2000) 065004
  [hep-th/9912162].

\bibitem{jjunct}
H.~Oda, K.~Ito, M.~Naganuma and N.~Sakai,
  Phys.\ Lett.\ B {\bf 471} (1999) 140
  [hep-th/9910095];
 K.~Ito, M.~Naganuma, H.~Oda and N.~Sakai,
  Nucl.\ Phys.\ B {\bf 586} (2000) 231
  [hep-th/0004188];
 M.~Naganuma, M.~Nitta and N.~Sakai,
  Phys.\ Rev.\ D {\bf 65} (2002) 045016
  [hep-th/0108179];
  K.~Kakimoto and N.~Sakai,
  Phys.\ Rev.\ D {\bf 68} (2003) 065005
  [hep-th/0306077].

\bibitem{jwallweb}
M.~Eto, Y.~Isozumi, M.~Nitta, K.~Ohashi and N.~Sakai,
  Phys.\ Rev.\ D {\bf 72} (2005) 085004
  [hep-th/0506135];
 M.~Eto, Y.~Isozumi, M.~Nitta, K.~Ohashi and N.~Sakai,
  Phys.\ Lett.\ B {\bf 632} (2006) 384
  [hep-th/0508241].

\bibitem{SY-abw}
  M.~Shifman and A.~Yung,
  Phys.\ Rev.\ D {\bf 67} (2003) 125007
  [hep-th/0212293].

\bibitem{j-boojums}
  Y.~Isozumi, M.~Nitta, K.~Ohashi and N.~Sakai,
  Phys.\ Rev.\ D {\bf 71} (2005) 065018
  [hep-th/0405129].

\bibitem{SY-naw}
 M.~Shifman and A.~Yung,
  Phys.\ Rev.\ D {\bf 70}, 025013 (2004)
  [hep-th/0312257].

  \bibitem{rev1}
  N.~Sakai and D.~Tong,
  JHEP {\bf 0503}, 019 (2005)
  [hep-th/0501207].

\bibitem{ASY-boojums}
  R.~Auzzi, M.~Shifman and A.~Yung,
  Phys.\ Rev.\ D {\bf 72} (2005) 025002
  [hep-th/0504148].

\bibitem{SY-mon}
M.~Shifman and A.~Yung,
  Phys.\ Rev.\ D {\bf 70}, 045004 (2004)
  [hep-th/0403149].

\bibitem{HT2}
A.~Hanany and D.~Tong,
JHEP {\bf 0404} (2004) 066,
[hep-th/0403158].

\bibitem{012}
  R.~Auzzi, S.~Bolognesi and J.~Evslin,
  JHEP {\bf 0502}, 046 (2005)
  [hep-th/0411074].

\bibitem{rev2}
D.~Tong,
{\em TASI Lectures on Solitons,}
 hep-th/0509216.

\bibitem{Jrev}
M.~Eto, Y.~Isozumi, M.~Nitta, K.~Ohashi and N.~Sakai,
{\em Solitons in the Higgs Phase --- the Moduli Matrix Approach},
hep-th/0602170.

\bibitem{skyrmioistantoni}
  M.~Eto, M.~Nitta, K.~Ohashi and D.~Tong,
  Phys.\ Rev.\ Lett.\  {\bf 95} (2005) 252003
  [arXiv:hep-th/0508130].

\bibitem{polyakov}
A.~M.~Polyakov,
  Nucl.\ Phys.\ B {\bf 120}, 429 (1977).

\bibitem{DS}
G.~R.~Dvali and M.~A.~Shifman,
  Phys.\ Lett.\ B {\bf 396}, 64 (1997)
  (E)  {\bf  B407}, 452 (1997),
  [hep-th/9612128].

  \bibitem{sm}
   A.~Kovner, M.~A.~Shifman and A.~Smilga,
  Phys.\ Rev.\ D {\bf 56}, 7978 (1997)
  [hep-th/9706089].

\bibitem{vafa}
  B.~S.~Acharya and C.~Vafa,
{\em On Domain Walls of ${\cal N} = 1$ Supersymmetric Yang--Mills in Four Dimensions,}
hep-th/0103011.

 \bibitem{ritz1}
  A.~Ritz, M.~Shifman and A.~Vainshtein,
  Phys.\ Rev.\ D {\bf 66}, 065015 (2002)
  [hep-th/0205083];

   \bibitem{ritz2}
  A.~Ritz, M.~Shifman and A.~Vainshtein,
  Phys.\ Rev.\ D {\bf 70} (2004) 095003
  [hep-th/0405175].

\bibitem{tong-mon}
D.~Tong,
Phys.\ Rev.\ D {\bf 69} (2004) 065003,
[hep-th/0307302].

\bibitem{n=1*}
  V.~Markov, A.~Marshakov and A.~Yung,
  Nucl.\ Phys.\ B {\bf 709}, 267 (2005)
  [hep-th/0408235].

   \bibitem{HSZ}
A.~Hanany, M.~J.~Strassler and A.~Zaffaroni,
  Nucl.\ Phys.\ B {\bf 513}, 87 (1998)
  [hep-th/9707244].

\bibitem{VY}
 A.~I.~Vainshtein and A.~Yung,
  Nucl.\ Phys.\ B {\bf 614}, 3 (2001)
  [hep-th/0012250].

\bibitem{tsm}
  J.~P.~Gauntlett, D.~Tong and P.~K.~Townsend,
  Phys.\ Rev.\ D {\bf 63} (2001) 085001
  [hep-th/0007124];
  Phys.\ Rev.\ D {\bf 64} (2001) 025010
  [hep-th/0012178];
  J.~P.~Gauntlett, R.~Portugues, D.~Tong and P.~K.~Townsend,
  Phys.\ Rev.\ D {\bf 63} (2001) 085002
  [hep-th/0008221].

\bibitem{jsm}
  M.~Arai, M.~Naganuma, M.~Nitta and N.~Sakai,
  Nucl.\ Phys.\ B {\bf 652} (2003) 35
  [hep-th/0211103];
{\em BPS Wall in ${\cal N} = 2$ SUSY Nonlinear Sigma Model with Eguchi-Hanson Manifold,}
in {\sl Garden of Quanta}, Eds. J. Arafune, et al., (World Scientific, Singapore, 2003),
p. 299, [hep-th/0302028].

\bibitem{ANO}
A.~Abrikosov, Sov.~Phys. JETP {\bf32}, 1442  (1957)
[Reprinted in {\em Solitons and Particles}, Eds. C. Rebbi and G. Soliani
(World Scientific, Singapore, 1984), p. 356];\\
H.~Nielsen and P.~Olesen, Nucl.~Phys. {\bf B61}, 45 (1973)
[Reprinted in {\em Solitons and Particles}, Eds. C. Rebbi and G. Soliani
(World Scientific, Singapore, 1984), p. 365].

\bibitem{Tbrane}
  D.~Tong,
JHEP {\bf 0602}, 030 (2006)
  [hep-th/0512192];

\bibitem{SYdual}
M.~Shifman and A.~Yung,
{\em Bulk-Brane Duality in Field Theory,}
 hep-th/0603236.

\bibitem{Bo}
E. Bogomolny,
  Sov. \ J. \ Nucl. \ Phys.  {\bf 24},  449 (1976).
[Reprinted in {\em Solitons and Particles}, Eds. C. Rebbi and G. Soliani
(World Scientific, Singapore, 1984), p. 389].




\end{thebibliography}
\end{document}